\documentclass{aa}
\usepackage{graphics}
\usepackage{psfig}

\newcommand{\nai}{Na{\sc i}}

\newcommand{\nii}{N{\sc ii}}

\newcommand{\hi}{H\,{\sc i}}

\newcommand{\etal}{et al.}

\newcommand{\Ein}{{\em Einstein}}

\newcommand{\ros}{{\em ROSAT}}
\newcommand{\Ros}{{\em ROSAT}}
\newcommand{\ltsim}{\raisebox{-1mm}{$\stackrel{<}{\sim}$}}
\newcommand{\gtsim}{\raisebox{-1mm}{$\stackrel{>}{\sim}$}}

\newcommand{\nh}{$N_{\rm H}$}

\newcommand{\eg}{e.g. }
\newcommand{\ie}{i.e. }

\def\arcm{\hbox{$^\prime$}}

\begin{document}

   \thesaurus{11;(11.01.2;
		11.09.1 IC~4329A; 
		11.09.2; 
		11.09.3;
		11.19.1; 
		13.25.2)}

   \title{ROSAT observations of the IC~4329A galaxy group\thanks{
	based partially on observations collected at the European Southern Observatory, 
 	La Silla, Chile}}
   \author{A.M.~Read \and W.~Pietsch}
   \offprints{A.M.\,Read, e-mail address aread@mpe.mpg.de}
   \institute{Max-Planck-Institut f\"ur extraterrestrische Physik,
              Gie\ss enbachstra\ss e, D--85748 Garching, Germany}
   \date{Received date April 27 1998; accepted date May 14 1998}
   \maketitle
   \markboth{A.M.~Read \& W.~Pietsch: ROSAT observations of the IC~4329A galaxy group}{}

\begin{abstract}

We report here the results of high-resolution and spectral imaging X-ray observations,
with both the \ros\ HRI and PSPC, of the field surrounding the nearby ($D=64$\,Mpc)
type~1 Seyfert galaxy IC~4329A and its giant lenticular companion IC~4329. Many point
sources are detected, the brightest being associated with IC~4329A itself, having an
extremely bright X-ray luminosity of $6\times 10^{43}$\,erg s$^{-1}$, and spectral
properties compatible with a single power-law model ($\Gamma=1.73$), with a spectral
break at 0.7\,keV. Two other bright sources are detected associated with the companion
galaxy IC~4329, and a likely quasar 14\arcm\ to the south-west. We have also 
established, through optical observations taken at the European Southern Observatory, 
that three further X-ray point sources, intriguingly positioned with respect to IC~4329A,
are in fact nothing to do with the system, and are merely foreground and background
objects.

In addition to point
source emission, residual, unresolved emission is detected surrounding the
IC~4329A/ IC~4329 pair, extending for some 200\,kpc. This emission appears markedly
two-component, comprising of a spectrally hard and smooth component,
circularly-distributed about the central galaxy pair, and a spectrally soft, more clumpy
component, positioned almost entirely to the south-east of IC~4329A. The hard component
of the residual emission itself appears two-component, one component being due to the
`wings' of the intensely bright IC~4329A source, the other, apparently due to hot
($\sim1.5$\,keV) gas, likely associated with the galaxy group of which IC~4329A and
IC~4329 are members. The soft component of the residual emission may be a larger
version of the superwinds seen around some ultraluminous far-infrared galaxies, or may
even represent a `stripped wake' of intragroup gas. Evidence for shocked gas due to the
central IC~4329A/ IC4329 interaction is also found between the two central galaxies.


\end{abstract}

\section{Introduction}
\label{sec_intro}

IC~4329A is a nearby ($z=0.01605$), edge-on, S0, type 1 Seyfert galaxy, at a
distance of 64\,Mpc  ($H_{0} = 75$\,km s$^{-1}$ Mpc$^{-1}$), situated very close
to the centre of Abell cluster A3574. This is unusual
given that Seyfert galaxies, like most spirals, are rarely found inside
clusters (\eg Osterbrock \cite{Osterbrock}). This cluster, first described by Shapley
(\cite{Shapley}), is often referred to as the IC~4329 cluster or group, after its
brightest member (\eg de Vaucouleurs \etal\ \cite{RC2}), or as Klemola~27
(Klemola  \cite{Klemola}), and is the easternmost and most distant member of the chainlike
Hydra-Centaurus Supercluster (Chincarini \& Rood \cite{Chincarini}).

The heavily reddened type~1 Seyfert nucleus observed within IC~4329A (Winkler
\etal\ \cite{Winkler}) is a strong FIR/X-ray source, the 25$\mu m$ peak in the {\em
IRAS} flux density, suggestive of there being hot gas close to the
nucleus. It has the steepest Balmer decrement ($H{\alpha}/H{\beta}\simeq 12$)
of any known Seyfert, a very steep optical spectral index ($\alpha = 4.4$),
and it is seen to show \nai\ in absorption (Penston \& Wilson \cite{Penston}). Although calculated
extinction values indicate that the absolute visual magnitude $M_{v}$, is at least -23,
and may be as high as -25, IC~4329A appears optically, to be a largely
undisturbed edge-on system, with a prominent dust lane.

The brightest galaxy within the cluster, the giant lenticular IC~4329 has a redshift
close to that of IC~4329A ($\Delta v = 460$\,km s$^{-1}$), and lies at a projected
distance of 59\,kpc to the west. As pointed out by Kollatschny \& Fricke (\cite{Kollatschny}),
these two systems appear to be part of a loose group of seven galaxies, indicating
that a number of the group member's activity may well be linked to past
interactions. The fact that IC~4329 is a shell galaxy, together with the fact low
surface brightness features are observed around IC~4329A (Wolstencroft \etal\
\cite{Wolstencroft}), suggest that an interaction is taking place between IC~4329 and IC~4329A.
Of the seven group members, only IC~4329A shows any indication of significant
nuclear activity.

Further evidence for possibly interaction-induced activity is also apparent
from emission-line imaging studies. A halo of emission-line gas can be seen 
around IC~4329A in
the $H{\alpha} +$[\nii] image of Colbert \etal\ (\cite{Colbert}), extending along the
minor axis, $\sim$10\arcsec\ (3\,kpc) on both sides of the nucleus, with a
luminosity of $\sim2.5\times10^{39}$erg s$^{-1}$. An almost identical image is
seen in Mulchaey \etal\ (\cite{Mulchaey96}). This $H{\alpha}$ halo is
believed by both sets of authors to represent an outflow from the nucleus,
perhaps a superwind of the type commonly seen in edge-on infrared-luminous
galaxies (\eg McCarthey \etal\ \cite{McCarthey}; Armus \etal\
\cite{Armus}).

Radio mapping of the IC~4329A system has brought about some very
interesting results also. In a Molonglo Synthesis Telescope (MOST) 843\,MHz survey
of radio sources in southern Abell clusters, IC~4329A is seen to
possess an apparent radio tail $\sim6$\arcm\ long (Unewisse \cite{Unewisse}),
corresponding to a linear size of 110\,kpc. This feature implies a radio
structure larger than any other known Seyfert, though it must be stressed that
the resolution of the MOST observations make it impossible to be certain
whether the observed feature is associated with the galaxy or not. Other significant
radio observations include those of Unger \etal\ (\cite{Unger}) at 1490\,MHz (and 
4860\,MHz), where, besides a bright core being visible, an extended region of
emission is observed extending $\sim6$\arcsec\ towards IC~4329. This feature may
well be associated with the interaction. Furthermore, Blank \& Norris (\cite{Blank}),
in a 4\arcsec\ resolution 2.3\,GHz observation, observed two 10\arcsec\ extensions
emanating from the nucleus. The first is roughly in the same direction as the
extension seen by Unger \etal\ (\cite{Unger}). The second, extending to the south-east, 
perpendicular to the dust-lane of the galaxy, may, according to Blank \& Norris
(\cite{Blank}), be due to a superwind, driven from the nucleus. Baum \etal\ (\cite{Baum})
found that the kpc-scale radio emission seen in several Seyfert galaxies
tended to align itself with the systems' minor axes, and they concluded that 
circumnuclear starbursts are the most likely cause. 

\subsection{Previous X-ray observations}

X-ray emission associated with IC~4329A was first suspected when Ariel~V detected the
steady source 2A~1347-300 (Cooke \etal\ \cite{Cooke}). The error box however, contained IC~4329
as well, and it was only later (Delvaille \cite{Delvaille}) that the source was identified with
IC~4329A. Further observations were performed by HEAO~1, both in scanning (Piccinotti
\etal\ \cite{Piccinotti}) and pointed modes (Tennant \& Mushotzky \cite{Tennant}; Mushotzsky 
\cite{Mushotzky84}), and by
HEAO~2 (Petre \etal\ \cite{Petre}). IC~4329A and IC~4329 also appear in Fabbiano \etal's 
(\cite{Fabbiano92}) X-ray catalog and atlas.

IC~4329A has also been observed (Miyoshi \etal\ \cite{Miyoshi}) with gas scintillation
proportional counters aboard the Japanese satellite Tenma (Tanaka \cite{Tanaka}). It had been
established prior to this that large amplitude variabilities in the X-ray flux are
rare within Seyfert galaxies on timescales of seconds to years (Mushotzky \cite{Mushotzky84}), and
indeed, no significant change in flux or spectral shape was seen over the six day
Tenma observation. The spectral shape above 15\,keV however, an almost flat, rather
positively-sloped tail (rare for Seyfert~1 galaxies), appeared to be very different
when compared to the HEAO~1 data (Mushotzky \cite{Mushotzky84}). Observations of IC~4329A with {\em
Ginga} (Piro \etal\ \cite{Piro}) showed a hard X-ray bump above 8\,keV,
probably produced by absorption or reflection of the central emission by a very thick
cold medium close to the nucleus, an idea supported by the detection of 6.4\,keV
fluorescence lines.

The \Ros\ PSPC data from IC~4329A have been published recently in conjunction with
the {\em COMPTON GRO} observations (Madejski \etal\ \cite{Madejski}, hereafter M95), their main
result being that IC~4329A's spectrum is compatible with a single power law, of energy
spectral index 1, modified by absorption and reflection extending from soft X-rays to
$\gamma$-rays. What evidence there is of a $\gamma$-ray spectral break is weak, and in
any case, the energy of this possible break is both higher than that of NGC~4151
(Zdziarski \etal\ \cite{Zdziarski}), and higher than that of typical Seyfert 1s (Fabian \etal\
\cite{Fabian}). The \Ros\ spectrum of IC~4329 was also presented, it being well described by an
optically thin thermal plasma with $kT=0.9$\,keV. As far as the spatial structure
obtained with the PSPC is concerned, M95 describe IC~4329A's X-ray source
as being consistent with point-like.

Lastly, {\em ASCA} observations of IC~4329A (Cappi \etal\ \cite{Cappi}) indicate that the
0.4$-$10\,keV spectrum is best described by a steep power law spectrum passing through
a warm absorber, together with a strong reflection component and Fe K line, confirming
both the above {\em ROSAT-GRO} (M95) and separate {\em ASCA}
(Mushotzky \etal\ \cite{Mushotzky95}) results. Furthermore, as concluded by M95, 
cold absorption in excess of the Galactic value is required by the data, consistent
with the edge-on nature of the galactic disc. 

The \Ros\ X-ray telescope (XRT), with the Position Sensitive Proportional
Counter (PSPC) (Pfeffermann \etal\ \cite{Pfeffermann}) at its focal plane, offers three
very important improvements over previous X-ray imaging instruments (such as
the \Ein\ IPC). Firstly, the spatial resolution is very much
improved, the 90\% enclosed energy radius at 1\,keV being $27''$
(Hasinger \etal\ \cite{Hasinger}). Secondly, the PSPC's spectral resolution is
very much better ($\Delta E/E \sim 0.4$ FWHM at 1\,keV) than earlier
X-ray imaging instruments, allowing the derivation of characteristic
source and diffuse emission temperatures. Lastly, the PSPC internal
background is very low ($\sim3\times10^{-5}$\,ct s$^{-1}$ arcmin$^{-2}$;
Snowden \etal\ \cite{Snowden}), thus allowing the mapping of low surface brightness
emission. The High Resolution Imager (HRI) on the other hand, because of
its excellent spatial resolution (more like $5''$) and relative insensitivity to
diffuse emission, is an ideal instrument for further investigation into the point
source populations (see Tr\"{u}mper (\cite{Trumper}) for a description of the \Ros\ satellite
and instruments).

Here we report the results of a 15.5\,ks \Ros\ High Resolution Imager (HRI)
observation that addresses one essential aspect of the X-ray emission from IC~4329A
and its neighbours that has not been possible until now $-$ that of the high
resolution spatial properties of the X-ray emission. Although, as mentioned above, some
aspects of the PSPC data have been published (M95), the authors
concentrate almost entirely on the spectral properties of the individual systems, 
and no discussion of the spatial properties is given. We have therefore
performed a thorough reanaysis 
of the 8.3\,ks of \Ros\ PSPC data, concentrating on the spatial
properties and on the existence of any extended features. These results are also
presented here.

The plan of the paper is as follows. Sect.\,\ref{sec_obs_data} describes the
observation and the preliminary data reduction methods used, 
Sect.\,\ref{sec_disc_sources} discusses the results as regards the point source
emission, Sect.\,\ref{sec_disc_unres} discusses the results as regards the remaining
unresolved emission, and finally a summary is presented in Sect.\,\ref{sec_summary}.

\section{ROSAT observations and preliminary analysis}
\label{sec_obs_data}

\subsection{HRI data}
\label{sec_hri}

The field surrounding IC~4329A was observed with the \Ros\ HRI on the 14th of
January 1995 in essentially two separate observations for a total 15.5\,ks. The
data appeared essentially very clean, and only 200 seconds of data were removed on
the basis of very high ($>20$\,ct s$^{-1}$) and very low ($<1$\,ct s$^{-1}$) count
rates, and on large values of atomic oxygen column density along the line of sight.

Source detection and position determination were performed over
the full field of view ($\sim37$\arcm\ diameter) with the EXSAS local detect, map
detect, and maximum likelihood algorithms (Zimmermann \etal\ \cite{Zimmermann92}). 
Images of pixel
size 5\arcsec\ were used for the source detection. During the source detection
procedure, only events detected in the raw HRI channels 2$-$8 were used, thus
reducing the UV/cosmic ray background. 

Sources accepted as detections were those with a likelihood L $\geq$8, and this gave
rise to a number of sources. Probabilities P, are related to maximum likelihood
values L, by the relation P$=1-e^{-\mbox{L}}$. Thus a likelihood L of 8 corresponds
to a Gaussian significance of 3.6$\sigma$ (Cruddace \etal\ \cite{Cruddace}; 
Zimmermann \etal\
\cite{Zimmermann94}). The source searching algorithms had difficulty in reliably detecting sources
close to the bright central feature. Sources that were `detected' here generally
appeared to be extended (\ie they had a large value of extension likelihood), while
possessing only a small number of counts. This is very suggestive of them being
merely spurious features situated in the wings of the bright central feature. These
sources were excluded from the analysis, leaving the 17 listed in
Table~\ref{table_srclist_hri}.

In an effort to improve the accuracy of the HRI source positions, the RA and Dec of
five bright sources (H2, H7, H15, H16 and H17) were compared with the APM finding
charts of Irwin \etal\ (\cite{Irwin}). The sources associated with the two central
galaxies IC~4329A (H11) and IC~4329 (H5), although they appeared to be very
coincident, were not used in the re-aligning process, because of the extended nature
of both their optical and X-ray emission. A very small offset of 0.10\arcsec\ in right
ascension and 2.1\arcsec\ in declination is observed.

Table~\ref{table_srclist_hri} lists the 17 detected HRI sources as follows:
source number (col.\,1, prefixed by a `H' to distinguish from PSPC-detected
sources, see Sect.\,\ref{sec_pspc}), corrected right ascension and declination
(cols\,.2 and 3), error on the source position (col.\,4, including a 3.9\arcsec\
systematic attitude solution error), likelihood of existence (col.\,5), net
counts and error (col.\,6) count rates and errors after applying deadtime and
vignetting corrections (col.\,7), and 0.1$-$2.4\,keV flux, assuming a 5\,keV
thermal bremsstrahlung model (col.\,8). Count rates of the HRI-detected point
sources can be converted into fluxes (and into luminosities), assuming other
additional spectral models, \eg\ thermal bremsstrahlung and Raymond \& Smith
hot plasma models (with cosmic abundances) at temperatures of 0.3\,keV and
3.0\,keV. HRI Conversion factors (in units of $10^{-11}$\,erg cm$^{-2}$
cts$^{-1}$ s$^{-1}$) for these models, from count rate (in s$^{-1}$) into flux
(corrected for Galactic absorption), are as follows: Thermal bremsstrahlung:
11.4 (0.3\,keV) and 5.3 (3.0\,keV); Raymond \& Smith hot plasma: 11.0
(0.3\,keV) and 5.2 (3.0\,keV). As a final point, only source H5 is flagged as
extended (at a likelihood of 173), with a FWHM of 12.4\arcsec.

\begin{table*}
\caption[]{X-ray properties of point sources detected with the HRI (see text).
Tabulated fluxes assume a 5\,keV thermal bremsstrahlung model and a hydrogen column
density of $N_{\rm H} = 4.4 \times 10^{20}$~cm$^{-2}$ (see text for conversion factors
for different temperatures/models).}
\label{table_srclist_hri}
\begin{tabular}{lrrrrrrr}
\hline
\noalign{\smallskip}
Source & R.A. (J2000) & Dec. (J2000) & R$_{err}$ & Lik. & Net counts & Count rate &
$F_{\rm x}$ \\
& ($^{\rm h~~~~m~~~~s}$) & ($^{\circ}~~~~\arcm~~~~\arcsec$) & ($\arcsec$) &  & &
($10^{-4}$\,s$^{-1}$) & ($10^{-14}\frac{{\rm erg}}{{\rm cm}^{2} {\rm s}}$) \\
\noalign{\medskip}
(1) & (2) & (3) & (4) & (5) & (6) & (7) & (8) \\
\noalign{\smallskip}
\hline
\noalign{\smallskip}
H1 & 13 48 44.5 & -30 29 47 & 4.6 & 354.3  & 256.4$\pm$17.4  & 183.2$\pm$12.4  & 95.4$\pm$6.5\\ 
H2 & 13 48 57.8 & -30 22 05 & 4.2 & 128.6  & 58.8$\pm$8.0    & 39.6$\pm$5.4    & 20.6$\pm$2.9 \\ 
H3 & 13 48 58.2 & -30 18 30 & 5.5 & 8.4    & 7.6$\pm$3.2     & 5.1$\pm$2.1     & 2.7$\pm$1.1 \\ 
H4 & 13 48 59.5 & -30 22 10 & 4.3 & 81.4   & 45.1$\pm$7.2    & 30.3$\pm$4.8    & 15.8$\pm$2.5 \\ 
H5 & 13 49 05.2 & -30 17 43 & 4.1 & 314.2  & 323.7$\pm$19.3  & 215.7$\pm$12.9  & 112.3$\pm$6.6 \\ 
H6 & 13 49 06.5 & -30 23 10 & 5.0 & 15.0   & 13.8$\pm$4.2    & 9.3$\pm$2.8     & 4.9$\pm$1.5 \\ 
H7 & 13 49 10.1 & -30 20 46 & 4.8 & 13.2   & 9.9$\pm$3.4     & 6.6$\pm$2.3     & 3.4$\pm$1.2 \\ 
H8 & 13 49 14.3 & -30 20 02 & 5.3 & 8.6    & 8.4$\pm$3.4     & 5.6$\pm$2.2     & 2.9$\pm$1.2 \\ 
H9 & 13 49 15.7 & -30 18 23 & 8.1 & 14.8   & 22.1$\pm$5.8    & 14.7$\pm$3.8    & 7.7$\pm$2.0 \\ 
H10& 13 49 18.8 & -30 17 30 & 8.2 & 11.1   & 16.1$\pm$5.0    & 10.7$\pm$3.3    & 5.6$\pm$1.7 \\ 
H11& 13 49 19.2 & -30 18 34 & 3.9 & 23751.3&10577.9$\pm$103.1&7003.2$\pm$68.3& 3645.0$\pm$35.6 \\ 
H12& 13 49 20.9 & -30 19 01 & 8.6 & 12.8   & 28.9$\pm$6.8    & 19.1$\pm$4.5    & 9.9$\pm$2.3 \\ 
H13& 13 49 21.6 & -30 18 34 & 9.7 & 10.3   & 30.3$\pm$7.2    & 20.0$\pm$4.8    & 10.4$\pm$2.5 \\
H14& 13 49 24.0 & -30 25 43 & 6.3 & 8.8    & 10.2$\pm$3.8    & 6.9$\pm$2.6     & 3.6$\pm$1.4 \\ 
H15& 13 49 39.2 & -30 30 17 & 8.1 & 26.5   & 39.7$\pm$8.0    & 28.0$\pm$5.6    & 14.6$\pm$3.0 \\ 
H16& 13 49 39.4 & -30 16 23 & 4.2 &108.4   & 46.0$\pm$7.6    & 35.1$\pm$5.1    & 18.2$\pm$2.7 \\ 
H17& 13 50 03.7 & -30 08 33 & 11.3& 21.2   & 46.0$\pm$9.5    & 33.1$\pm$6.8    & 17.2$\pm$3.5 \\ 
\noalign{\smallskip}
\hline
\end{tabular}
\end{table*}

Fig.\,1 shows the inner 25$\times$25 arcminute field of view (where all of
the 17 sources listed in Table~\ref{table_srclist_hri} are visible). Sources
coincident with both IC~4329A (H11, with sources H9, H10, H12 and H13 in close
proximity) and IC~4329 (H5) are visible, as is a further bright source (H1),
15\arcm\ to the southwest of IC~4329A. 

\begin{figure*}
\vspace{75mm}
\hfill \parbox[b]{5.5cm} 
{\caption{Grey-scale image of the X-ray flux seen with the \Ros\ HRI over the central
25$\times$25\arcm\ field. The image was constructed with a pixel size of 2\arcsec, 
and smoothed with a Gaussian of FWHM 10\arcsec. The point sources, as given in 
Table~\ref{table_srclist_hri}, are enclosed by boxes and numbered.}}
\end{figure*}

To investigate whether any residual, low
surface brightness emission is visible, an adaptive filtering technique was used to
create an intensity-dependent smoothed image. The signal-to-noise level of low
surface brightness emission is boosted by smoothing lower and lower intensity
sections of the image using Gaussians of progressively larger FWHM. Photons (again,
from raw channels 2$-$8) were binned into an image of pixel size 8\arcsec. Pixels of
amplitude 1 (2,3,4,5,6,7,8) were smoothed with a Gaussian of  FWHM 170\arcsec
(120,85,60,40,30,20,15)\arcsec. Pixels of amplitude greater than 8 remained
unsmoothed, thus ensuring that the bright point sources were not smoothed into the
background regions. The resultant image is shown as a contour map overlayed on an 
optical image in Fig.\,2 (the optical image is taken from the U.K.\,Schmidt plate 
digitised sky survey).

\begin{figure*}
\vspace{75mm}
\hfill \parbox[b]{5.5cm} 
{\caption{\Ros\ HRI map of the IC~4329A field obtained using an adaptive filtering
technique (see text), overlayed on a digitized sky survey image. 
Only channels 2$-$8 are used. 
The contour levels are at 2, 3,
5, 9, 15, 31, 63, 127, 255, 511, 1023 and 2047$\sigma$ ($\sigma$ being
$1.0\times10^{-3}$\,cts s$^{-1}$ arcmin$^{-2}$) above the background
($12.9\times10^{-3}$\,cts s$^{-1}$ arcmin$^{-2}$). }}
\end{figure*}

\subsection{PSPC data}
\label{sec_pspc}

The field surrounding IC~4329A was observed with the \Ros\ PSPC for a total 
of 8.23\,ks, in five separate observation intervals between the 12th and the 14th 
of January 1993. 

Source detection and position determination were performed over the full field of view as
with the HRI data, using the EXSAS local detect, map detect, and maximum likelihood
algorithms (Zimmermann \etal\ \cite{Zimmermann92}). As with the HRI analysis, it is very
doubtful whether many of the sources initially detected close to the bright central source,
associated with IC~4329A, are true detections. Again, as with the HRI, those sources close
to the bright central source with a large value of extension likelihood and a low number of
counts were excluded from the analysis, leaving 22 sources.

As with the HRI analysis, an effort was made to correct the attitude solution.
Comparisons were made between the positions of 4 bright point sources (P3, P11, P12,
P14) and the APM finding charts (Irwin \etal\ \cite{Irwin}). The positions of the
bright sources associated with the central galaxies were not used in the
re-aligning, on account of their extended nature. An offset (purely in the
north-south direction) of 7\arcsec\ is seen.

22 sources were detected with a likelihood L $\geq$10 over the entire PSPC field of
view. The 17 sources detected within the central 30\arcm$\times$30\arcm\ area are
listed below in Table~\ref{table_srclist_pspc} as follows: source number (col.\,1,
prefixed by a `P' for PSPC), corrected right ascension and declination
(cols.\,2,\,3), error on the source position (col.\,4, including a 3.9\arcsec\ 
systematic attitude solution error), likelihood of existence (col.\,5), net counts
and error (col.\,6) count rates and errors after applying deadtime and vignetting
corrections (col.\,7), and the (0.1$-$2.4\,keV) flux, assuming a 5\,keV thermal
bremsstrahlung model (col.\,8). The only source flagged as extended is P6 (corresponding to 
HRI source H5; see Table~\ref{table_srclist_hri}), at a likelihood of 186, and with 
a FWHM of 45.8\arcsec.

\begin{table*}
\caption[]{X-ray properties of point sources detected with the PSPC (see text).
Tabulated fluxes assume a 5\,keV thermal bremsstrahlung model and a hydrogen column
density of $N_{\rm H} = 4.4 \times 10^{20}$~cm$^{-2}$ (see text for conversion factors
for different temperatures/models). }
\label{table_srclist_pspc}
\begin{tabular}{lrrrrrrr}
\hline
\noalign{\smallskip}
Source & R.A. (J2000) & Dec. (J2000) & R$_{err}$ & Lik. & Net counts & Count rate &
$F_{\rm x}$ \\
 & ($^{\rm h~~~~m~~~~s}$) & ($^{\circ}~~~~\arcm~~~~\arcsec$) & ($\arcsec$) &  & &
($10^{-3}$\,s$^{-1}$) & ($10^{-14}\frac{{\rm erg}}{{\rm cm}^{2} {\rm s}}$) \\
\noalign{\medskip}
(1) & (2) & (3) & (4) & (5) & (6) & (7) & (8) \\
\noalign{\smallskip}
\hline
\noalign{\smallskip}
P1 &13 48 37.0&-30 13 10&12.3&20.2   & 26.6$\pm$6.5     &3.4$\pm$0.8    &   6.3$\pm$1.5  \\
P2 &13 48 42.9&-30 13 13&16.3&12.6   & 18.1$\pm$5.6     &2.3$\pm$0.7    &   4.3$\pm$1.3  \\
P3 &13 48 44.9&-30 29 42&4.6 &1767.4 & 697.6$\pm$27.3   &94.2$\pm$3.7   & 175.6$\pm$6.9  \\
P4 &13 48 58.6&-30 22 07&6.9 &151.9  & 111.4$\pm$11.7   &14.1$\pm$1.5   &  26.3$\pm$2.8  \\
P5 &13 49 03.0&-30 19 35&19.1&18.7   & 42.9$\pm$8.9     &5.4$\pm$1.1    &  10.1$\pm$2.1  \\
P6 &13 49 05.6&-30 17 46&5.1 &1273.4 & 647.2$\pm$26.7   &81.4$\pm$3.4   & 151.7$\pm$6.3  \\
P7 &13 49 07.0&-30 22 58&8.8 &37.7   & 35.4$\pm$7.0     &4.5$\pm$0.9    &   8.4$\pm$1.7  \\
P8 &13 49 19.4&-30 18 36&3.9 &26484.9& 21889.8$\pm$352.2&2746.4$\pm$44.2&5118.3$\pm$82.4 \\
P9 &13 49 28.6&-30 17 39&15.4&16.1   & 29.4$\pm$7.0     &3.7$\pm$0.9    &   6.9$\pm$1.7  \\
P10&13 49 30.4&-30 18 43&14.8&10.1   & 16.7$\pm$5.4     &2.1$\pm$0.7    &   3.9$\pm$1.3  \\
P11&13 49 38.9&-30 30 14&12.9&29.7   & 39.7$\pm$7.8     &5.3$\pm$1.0    &   9.9$\pm$1.9  \\
P12&13 49 39.4&-30 16 22&6.1 &158.3  & 100.0$\pm$10.9   &12.7$\pm$1.4   &  23.7$\pm$2.6  \\
P13&13 49 56.5&-30 10 30&17.3&10.4   & 17.8$\pm$5.6     &2.4$\pm$0.7    &   4.5$\pm$1.3  \\
P14&13 50 04.0&-30 08 39&10.4&58.2   & 60.2$\pm$9.1     &8.2$\pm$1.2    &  15.3$\pm$2.2  \\
P15&13 50 18.0&-30 07 54&21.5&13.1   & 25.7$\pm$6.9     &3.6$\pm$1.0    &   6.7$\pm$1.9  \\
P16&13 50 18.7&-30 26 06&15.9&21.6   & 35.1$\pm$7.7     &4.8$\pm$1.1    &   8.9$\pm$2.1  \\
P17&13 50 30.7&-30 10 43&23.7&11.8   & 25.8$\pm$7.2     &3.7$\pm$1.0    &   6.9$\pm$1.9  \\
\noalign{\smallskip}
\hline
\end{tabular}
\end{table*}

Fig.\,3 shows a broad band (channels 8$-$235, corresponding
approximately to 0.08$-$2.35\,keV) contour image of the central 
region close to the centre of the field of view.  All of the sources listed in
Table~\ref{table_srclist_pspc} are visible including sources associated with both
IC~4329A (P8) and IC~4329 (P6). Also shown in Fig.\,3 are three smaller
images, again showing the central emission, selected over three separate spectral
bands $-$ the `soft' band (channels 8$-$41), the `hard~1' band (channels 52$-$90)
and the `hard~2' band (channels 91$-$201). 

\begin{figure*}
\unitlength1.0cm 
\vspace{110mm}
\hfill \parbox[b]{18.0cm} 
{\caption{\Ros\ PSPC maps of the IC~4329A field in the broad (channels 8$-$235, 
corresponding approximately to 0.08$-$2.35\,keV) band
(main picture) and in the soft (channels 8$-$41), hard~1 (channels 52$-$90) and
hard~2 (channels 91$-$201) bands (three smaller pictures). The contour levels in
each figure are at 2, 3, 5, 9, 15, 31, 63, 127, 255, 511, 1023, 2047 and
4095$\sigma$ ($\sigma$ being $1.25\times10^{-3}$ (broad), $1.49\times10^{-3}$
(soft), $7.29\times10^{-4}$ (hard~1) and $1.42\times10^{-3}$ (hard~2) cts s$^{-1}$
arcmin$^{-2}$) above the background ($2.81\times10^{-3}$ (broad),
$3.40\times10^{-4}$ (soft), $5.41\times10^{-4}$ (hard~1) and $8.04\times10^{-4}$
(hard~2) cts s$^{-1}$ arcmin$^{-2}$). Source positions, as given in
Table~\ref{table_srclist_pspc}, are marked on the broad band image.
}}
\end{figure*}

Comparison of Fig.\,3 with Fig.\,1 shows many
sources visible in both the HRI and in the PSPC fields of view. 
Of the 9 HRI sources with a  high value ($>15$) of detection likelihood, {\em every
one} has a PSPC counterpart. The converse is also true; {\em all} 8 of the PSPC
sources with detection likelihoods greater than 27 (that lie within the HRI field
of view), have a HRI counterpart. We can be very confident therefore, that all of
these features are genuine point sources. The joint HRI/PSPC properties of these
bright sources, that we concentrate on below, are summarised in
Table~\ref{table_joint_sources} as follows: HRI source number (col.\,1), PSPC
source number (col.\,2), offset (in arcsec) between the (attitude-corrected) HRI
and PSPC positions (col.\,3), count rates and errors, after applying deadtime and
vignetting corrections, for the HRI (col.\,4), the PSPC (col.\,5), and the PSPC
analysis of M95 (col.\,6, - note that a different technique has been used here -
see M95). Fluxes, corrected for Galactic absorption, are given in col.\,7 (HRI) and
col.\,8 (PSPC) assuming in both cases a 5\,keV thermal bremsstrahlung spectrum. It
may appear that some discrepancy exists between the PSPC- and HRI-calculated fluxes,
especially in the brighter sources. This  could be due either to time variability
(discussed in the next section) or to the assumption of the wrong spectral model
(discussed in Sect.\,\ref{sec_disc_sources}). The final column (col.\,9) of
Table~\ref{table_joint_sources} gives the nearest bright optical counterpart, using
the APM finding charts of Irwin \etal\ (\cite{Irwin}); type (S$-$stellar, G$-$galaxy,
F$-$faint), B magnitude, and offset (from the HRI position) in arcsec.

\begin{table*}
\caption[]{X-ray properties of point sources detected with both the HRI and the PSPC
(see text). Tabulated fluxes for both the HRI and PSPC assume a 5\,keV thermal
bremsstrahlung spectrum, assuming a hydrogen column density of $N_{\rm H} = 4.4 \times
10^{20}$~cm$^{-2}$. Note that for H1-P3, H5-P6 and H11-P8, further detailed spectral
analysis has been performed (see Sect.\,\ref{sec_disc_sources}). Note also that the
PSPC source P4 is resolved by the HRI into two separate sources, H2 and H4. }
\label{table_joint_sources}
\begin{tabular}{llrrrrrrr}
\hline
\noalign{\smallskip}
HRI & PSPC & sep.\,(\arcsec) & \multicolumn{3}{c}{Count rate ($10^{-3}$\,s$^{-1}$)} & 
\multicolumn{2}{c}{$F_{\rm x}$ ($10^{-14}$\,erg cm$^{-2}$ s$^{-1}$)} &Identification  \\
ident. & ident.& & (HRI) & (PSPC) & (PSPC; M95) & (HRI) & (PSPC) & \\
\noalign{\medskip}
(1) & (2) & (3) & (4) & (5) & (6) & (7) & (8) & (9) \\
\noalign{\smallskip}
\hline
\noalign{\smallskip}
H1 &P3 &  7.2& 18.3$\pm$1.2 &  94.2$\pm$3.7 &  86.0$\pm$6.0 &     95.4$\pm$6.5 & 175.6$\pm$6.9  
	& S (16.5) 2.9\arcsec\ \\ \vspace{-0.9mm}
H2 &\raisebox{-1.4ex}{P4}& 10.6&  3.9$\pm$0.5 &\raisebox{-1.4ex}{14.1$\pm$1.5} &               
&     20.6$\pm$2.9 & \raisebox{-1.4ex}{26.3$\pm$2.8} & S (10.7) 1.9\arcsec\ 
\\ 
H4 &    &12.1&  3.0$\pm$0.5 &               &               &     15.8$\pm$2.5 &
	& F (20.8) 12.0\arcsec\ \\
H5 &P6 &  6.0& 21.6$\pm$1.3 &  81.4$\pm$3.4 &  81.0$\pm$4.0 &    112.3$\pm$6.6 & 151.7$\pm$6.3  
	&(G) IC~4329 \\
H6 &P7 & 13.6&  0.9$\pm$0.3 &   4.5$\pm$0.9 &               &      4.9$\pm$1.5 &  8.4$\pm$1.7
	& S (19.9) 3.4\arcsec\ \\
H11&P8&  3.3&700.3$\pm$6.8&2746.4$\pm$44.2&2650.0$\pm$20.0&   3645.0$\pm$35.6&5118.3$\pm$82.4
	& (G) IC~4329A\\
H15&P11&  4.9&  2.8$\pm$0.6 &   5.3$\pm$1.0 &               &     14.6$\pm$3.0 &  9.9$\pm$1.9  
	& S (18.9) 2.0\arcsec\ \\
H16&P12&  1.0&  3.5$\pm$0.5 &  12.7$\pm$1.4 &               &     18.2$\pm$2.7 & 23.7$\pm$2.6  
	& S (17.6) 1.9\arcsec\ \\
H17&P14&  7.1&  3.3$\pm$0.7 &   8.2$\pm$1.2 &               &     17.2$\pm$3.5 & 15.3$\pm$2.2  
	& S (14.7) 2.2\arcsec\ \\
\noalign{\smallskip}
\hline
\end{tabular}
\end{table*}

\subsection{Time variability study of point sources}
\label{timing}

A time variability study was performed for all of the sources detected in either the
ROSAT HRI and PSPC fields of view, with special attention paid to the nine brightest
sources (eight in the PSPC - see above) detected with both instruments.

For the HRI-detected sources, the complete observation was binned into five
observation blocks (of between 2 and 4\,ks). A maximum likelihood search at the
source positions given in Table~\ref{table_srclist_hri} was performed for each of
the five observation blocks, the vignetting and deadtime corrected count rates (and
errors) calculated within a cut radius of $1.5\times$ the PSF FWHM at the
source positions. Where a source was not detected with a likelihood ${\rm L} > 3.1$
(corresponding to a Gaussian significance of $2\sigma$), a $2\sigma$ upper limit to
the count rate was calculated. 
An essentially identical procedure was followed for the PSPC data, the observation
binned again into five observation blocks (of $\sim1.6$\,ks each). A cut radius of
$2.5\times$ the PSF FWHM was used.
For the very bright source IC~4329A, the timing analyses described above were
repeated for both the HRI and PSPC data, the observations being binned into 40
observation blocks.

Figs.~4 and 5 shows the results of this analysis for all the sources
(except for the bright IC~4329A source) detected with both the HRI and the PSPC
(Fig.\,4 shows the results for the HRI, Fig.\,5 for the PSPC). The layout of the
figures is such that the coincident sources appear at the same position (hence H2 and 
H4 being coincident with P6). The results of the timing analysis for IC~4329A are
shown in Fig.\,6 (both HRI and PSPC). 

Note that we are here interested in {\em variations} in the count rate  from these
sources, and not in the absolute values of these count rates, as these have been
calculated earlier (Table~3). Hence we have been able to use small cut radii to avoid
contamination from neighbouring sources and concentrate at the very centres of each 
source. Differences in the calculated count rates between these two methods appear to
be negligible, except in the HRI cases involving the very bright, extended sources,
where reductions in count rate, from those given in Table~3, can be seen. This is to 
be expected, given the methods involved.

It appears that little in the way of any temporal variability exists for any of the
bright sources (including that associated with IC~4329A), and this is borne out by fitting the light curves
to constant flux values. Lightcurve H2, in fact, shows the largest deviation from
consistency, a constant flux level fitting the data with a reduced $\chi^2$ of 2.6. The
H2 lightcurve therefore is variable only at the 2.1$\sigma$ significance level (or 
at likelihood of 3.42). Thus no significant variability is observed within any of the
detected sources.

\begin{figure*}
\vspace{15mm}
\hfill \parbox[b]{5.5cm} 
{\caption{\ros\ HRI lightcurves of the eight bright point sources detected with
both the HRI and the PSPC (excluding IC~4329A) from Table~\ref{table_joint_sources}.
Observation times are 1.5, 4.1, 4.1, 3.9, and 1.7\,ks.   
Count rates are shown by filled squares with 1$\sigma$ error bars, and 2$\sigma$ upper
limits are shown by open squares. Dashed lines indicate the average count rate
calculated over the complete observation.
}}
\end{figure*}

\begin{figure*}
\vspace{15mm}
\hfill \parbox[b]{5.5cm} 
{\caption{\ros\ PSPC lightcurves of the seven bright point sources detected with
both the HRI and the PSPC (excluding IC~4329A) from Table~\ref{table_joint_sources}.
Observation times are 1.6, 1.6, 1.6, 1.7, and 1.8\,ks.   
Count rates are shown by filled squares with 1$\sigma$ error bars, and 2$\sigma$ upper
limits are shown by open squares. Dashed lines indicate the average count rate
calculated over the complete observation.
}}
\end{figure*}

\begin{figure*}
%
\vspace{25mm}
\hfill \parbox[b]{5.5cm} 
{\caption{\ros\ HRI (top) and PSPC (bottom) lightcurves of IC~4329A.
The Observations were split into $\approx400$\,s (HRI) and $\approx200$\,s (PSPC)
bins. Count rates are shown by filled squares with 1$\sigma$ error bars, and
the dashed lines indicate the average
count rate calculated over the complete observation.
}}
\end{figure*}

\section{Results and discussion - the point sources}
\label{sec_disc_sources}

\subsection{The bright point sources - IC~4329A, IC~4329 \& H1-P3}

The three bright sources visible in the IC~4329A field are all especially
interesting. H1-P3 is bright, shows marginal evidence for extension in the
PSPC (though the fact that no evidence for extension is seen in the HRI data
implies that the source is truly unresolved) and appears associated with a
quite bright (B magnitude = 16.5) star-like object, some 2.9\arcsec\ distant. H5-P6 is
also very bright, appears to be significantly extended (both in the
PSPC and HRI data), and is associated with the giant elliptical galaxy
IC~4329. Finally, H11-P8 is extremely bright, showing a great deal of structure,
and is undoubtedly due to the Seyfert galaxy, IC~4329A.

It is worth noting again here that an analysis of the PSPC data (in
conjunction with {\em COMPTON GRO observations}) has already been published by
Madejski \etal\ (\cite{Madejski}) (M95). They deal however, almost exclusively with the
spectral properties of IC4329A (plus those of IC~4329 and H1-P3, or as they
call it, S3), and so, in the discussion that follows, many of the results we
present have not been addressed by M95, and are new, though we do compare 
the results of our spectral analysis with those of M95.

As in M95, spectra of all three bright objects (IC~4329A, IC~4329 and H1-P3)
were analysed. Spectra were extracted for all three
objects from within circles of 1.7\arcm\ (for IC~4329A and IC~4329) and
3\arcm\ (for H1-P3) at the position of each source (we note that an extraction
radius of 3\arcm\ for IC~4329A, as used in M95, is likely to be too large,
given that this is approximately the distance between IC~4329A and IC~4329).
Background spectra were extracted as follows: for IC~4329A, from
an annulus 6.6\arcm\ to 9.1\arcm\ from IC~4329A, thus avoiding contamination
from any other bright features; for IC~4329, from a 1.7\arcm\ radius circle
situated equidistant, on the opposite side of IC~4329A, thus ensuring that
contamination from the very bright central source could be removed; and for
H1-P3, from a 3\arcm\ to 5.5\arcm\ annulus centred on H1-P3, again avoiding
any bright features.

The three background-subtracted spectra, once corrected for exposure and
vignetting effects, were fitted with standard spectral models (thermal 
bremsstrahlung, power law, blackbody and Raymond \& Smith (\cite{Raymond}) hot plasma
models). A number of extra, more complex models have been attempted as regards
the IC~4329A spectrum, as in M95, and the results of all the best fits are 
given below in Table~\ref{table_fits_sources} as follows: Source (col.\,1),
spectral model (whether PL - power law plus absorption, BB - blackbody plus
absorption, PL/E - power law plus absorption and an edge, RS Raymond \& Smith
hot plasma plus absorption) (col.\,2), fitted $N_{\rm H}$ (col.\,3), fitted
spectral index $\Gamma$, where $F\propto E^{-\Gamma}$ (col.\,4), fitted
temperature (kT, in keV) (col.\,5) (Note here that in the case of the PL/E
model, this column gives the edge energy in keV). The next columns give the
metallicity  (solar, where an `F' indicates that the value has been frozen)
(col.\,6), the reduced $\chi^{2}$ (col.\,7), and three values of the
(0.1$-$2.4\,keV) luminosity (cols.\,8-10). Two values of luminosity as
calculated using the PSPC results are given; one (col.\,8) gives the `intrinsic'
luminosity of the source (\ie correcting for the total $N_{\rm H}$), the
second (col.\,9) gives an `emitted' luminosity (\ie correcting merely for the
Galactic $N_{\rm H}$). The final luminosity column (col.\,10) gives the
intrinsic (0.1$-$2.4\,keV) luminosity, using the count rate observed with the HRI,
and calculating the fluxes, assuming identical spectral models as inferred
from the PSPC data. All luminosities are calculated for an assumed distance of
64\,Mpc (which is almost certainly incorrect in the case of H1-P3, as
discussed below). 

\begin{table*}
\caption[]{Results of the best model fits to the IC~4329A, IC~4329 and H1-P3 spectra
(see text). Models are: PL (power law plus absorption), BB (blackbody plus
absorption), PL/E (power law plus absorption and an edge), RS (Raymond \& Smith hot
plasma plus absorption). In the case of the PL/E fit, the temperature $kT$ refers to the
temperature of the edge. Three (0.1$-$2.4\,keV) luminosities are tabulated. One, the
intrinsic PSPC luminosity of the source, two, the Galactic $N_{\rm H}$-corrected (\ie
emitted) PSPC luminosity (Galactic $N_{\rm H} = 4.4 \times 10^{20}$~cm$^{-2}$), and
lastly, the intrinsic HRI luminosity, using the HRI count rates in conjunction with
the models suggested by the PSPC data.}
\label{table_fits_sources}
\begin{tabular}{llrrrrrrrr}
\hline
\noalign{\smallskip}
Source & Model & $N_{\rm H}$ & Photon & $kT$  & $Z$     & red.$\chi^{2}$ & 
\multicolumn{3}{c} {$L_{\rm x}$ (10$^{42}$\,erg s$^{-1}$)} \\ 
 & & 10$^{20}$\,cm$^{-2}$& Index    &(keV)&(Solar)& & (Intrinsic)& (Emitted) & HRI (Intrinsic)\\
(1) & (2) & (3) & (4) & (5) & (6) & (7) & (8) & (9) & (10) \\
\noalign{\smallskip}
\hline
\noalign{\smallskip}
IC~4329A & PL  & 22.7$\pm$3.3 & 1.28$\pm$0.12 & &             & 1.1 & 41.6$\pm$0.7 & 23.7$\pm$0.4 & 29.0$\pm$0.3 \\
         & BB  &  7.5$\pm$0.1 & & 0.52$\pm$0.02 &             & 1.1 & 23.8$\pm$0.4 & 22.8$\pm$0.4 & 16.6$\pm$0.2 \\
         & PL/E& 27.9$\pm$5.1 & 1.73$\pm$0.33&(0.72$\pm$0.14)&& 0.9 & 63.0$\pm$1.0 & 23.4$\pm$0.4 & 43.0$\pm$0.5 \\
IC~4329  & RS  &  2.7$\pm$1.2 & & 1.08$\pm$0.06 & 1.0(F)      & 1.3 & 0.64$\pm$0.02& 0.64$\pm$0.02& 0.53$\pm$0.03 \\
         & RS  &  4.3$\pm$1.6 & & 1.07$\pm$1.07 &0.4$\pm$0.2  & 1.3 & 0.79$\pm$0.03& 0.79$\pm$0.03& 0.61$\pm$0.04 \\
H1-P3    & PL  &  3.2$\pm$1.3 & 2.35$\pm$0.33 & &             & 1.4 & 1.04$\pm$0.04& 1.04$\pm$0.04& 0.67$\pm$0.05 \\
\noalign{\smallskip}
\hline
\end{tabular}
\end{table*}

Although no thermal model (whether a Raymond \& Smith hot
plasma model or a thermal bremsstrahlung model) is able to fit the IC~4329A data adequately, a simple power law model does
gives quite an acceptable fit, the fitted parameters agreeing well with M95 and with Rush
\etal\ (\cite{Rush}). However, as in M95, close inspection of the residuals does suggest
an edge-like feature at around 0.7\,keV. Incorporating this edge into the model does
improve the fit (the data and residuals are shown in Fig.\,7), and we are able to
reproduce the best-fit 
results of M95 very accurately. Firstly, a photon index of
1.73$\pm$0.33 is suggested, consistent with M95, with the {\em Ginga} data (Piro \etal\
\cite{Piro}; Fiore \etal\ \cite{Fiore}), and with the {\em ASCA} data (Cappi \etal\
\cite{Cappi}). Secondly, the edge feature at 0.72$\pm$0.07\,keV is found at exactly the
same energy as in M95. As M95 suggest, the energy of this edge is inconsistent with that
expected if there were a neutral absorber present, and this strongly suggests the
presence of an ionized absorber (O{\sc vi}, O{\sc vii}). Further modelling, to address
the question of the true nature of this absorber, is possible. However, because of the
modest spectral resolution of the PSPC, one cannot distinguish between different models,
\ie between an ionized absorber model, a partial covering by neutral material model, and
a high column cold absorber model (see M95 for a detailed discussion). 

It is worth noting that the inferred $N_{\rm H}$, 27.9$\times10^{20}$\,cm$^{-2}$, is
substantially larger than the Galactic $N_{\rm H}$ in the direction of IC~4329A
(4.4$\times10^{20}$\,cm$^{-2}$; Dickey \& Lockman \cite{Dickey}),  indicating the
presence of a large intrinsic absorption. This is not too surprising given the edge-on
nature of the galaxy. In their study of the soft X-ray properties of Seyfert galaxies in
the \Ros\ All-Sky Survey, Rush \etal\ (\cite{Rush}) also find a very significant excess
in the best fit \nh.

The intrinsic (0.1$-$2.4\,keV) luminosity of IC~4329A, 6.3$\times10^{43}$\,erg
s$^{-1}$, is very large, within the top 10\% or so of the Seyferts within the Rush
\etal\ (\cite{Rush}) All-Sky Survey sample. It is an extremely luminous galaxy, 
and an extremely luminous Seyfert galaxy as well.

\begin{figure}
\unitlength1.0cm 
\label{fig_spec_ic4329a}
\vspace{120mm}
\hfill \parbox[b]{8.7cm} 
{\caption{IC~4329A spectrum with the best-fit power-law plus absorption edge model
(see Table~\ref{table_fits_sources}). The pulse height spectrum of the total X-ray 
emission is indicated by crosses, and the fit, by the solid line. 
}}
\end{figure}

The IC~4329 spectrum on the other hand, is only fitted adequately well by a thermal (Raymond \&
Smith hot plasma) spectrum. The best fit, while keeping the metallicity frozen at solar,
results in a well-constrained, 1.08$\pm$0.06\,keV spectrum, absorbed by a column of
2.7$(\pm1.2)$\\
$ \times 10^{20}$\,cm$^{-2}$, a column consistent with (though on the low side of)
the Galactic value. This result is entirely consistent with M95. Fitting of the spectrum while
letting the metallicity optimize, gives a column entirely consistent with the Galactic value
and a low (0.4$\pm$0.2 solar) metallicity, though the fitted temperature is less well
constrained than in the frozen-metallicity case. The intrinsic (0.1$-$2.4\,keV) luminosity of
IC~4329, 7.9$\times10^{41}$\,erg s$^{-1}$, is somewhat higher than average when compared to
optically similar systems (Fabbiano \etal\ \cite{Fabbiano92}). The fitted temperature is
entirely consistent with that of ellipticals, {\em ASCA} observations resulting in temperatures
for several early-type galaxies of between 0.7 and 1.2\,keV (Matsushita \etal\
\cite{Matsushita}; Rangarajan \etal\ \cite{Rangarajan}). Similarly, the low fitted metallicity
appears to be consistent with ellipticals, high-resolution studies with {\em ASCA} revealing
abundances $\leq0.5$\,solar in several cases (Loewenstein \etal\ \cite{Loewenstein}; Matsushita
\etal\ \cite{Matsushita}).

The H1-P3 spectrum is best fit with a power law model of photon index 2.35$\pm$0.33,
absorbed by a column of 3.2$(\pm$\\
$1.3) \times10^{20}$\,cm$^{-2}$, consistent with that
out of our own Galaxy. It is almost certainly due to a background quasar, given the
facts that it is unresolved in the HRI data, it appears coincident with a quite bright
(B mag = 16.5) star-like object, and it has a spectrum consistent with that of quasars
(power law with photon indices in the range 2.2$\pm$0.2; Branduardi-Raymont \etal
\cite{Branduardi}; Roche \etal\ \cite{Roche}).

Finally, note that, in Table~\ref{table_joint_sources}, it appeared that, in the cases
of these bright sources, the inferred HRI and PSPC fluxes did not agree particularly
well. This could have been attributable to time-variability or the assumption of the
wrong spectral model. We have already seen however (Figs.\,4$-$6) that none of these
sources appear to be particularly time-variable in either the HRI or the PSPC, and
usage of the correct spectral model, as has been done here, has only really aided the
situation in the case of IC~4329, and then, only slightly. In the case of H1-P3, it is
possible that the source has varied between the HRI and PSPC observations. The fact
that the object is very likely to be a QSO or background AGN adds some credence to
this. Also bare in mind that the quality of fit to the H1-P3 PSPC spectra is not
excellent, the reduced $\chi^{2}$ being only 1.4. In the case of IC~4329A and IC~4329,
the situation is rather intriguing. Both sources appear extended however, and this will 
lead to a reduction in the calculated HRI count rates, compared to the PSPC count rates.
Furthermore, IC~4329A is extremely bright, accentuating the above effect. Finally, as
may be the case (Fig.\.2), and is discussed in detail later, if there were a large
amount of low-surface brightness, diffuse emission in the vicinity of these two
galaxies, this could very well lead to a reduction in the HRI count rates compared to
the PSPC count rates, the HRI being relatively far less sensitive to this type of
emission than the PSPC. 

\subsection{The secondary point sources}

Moving on to the remaining point sources, many interesting results have been obtained.
Feature P4, for instance, appears elongated in the east-west direction in the PSPC
image (Fig.\,3). The HRI is able to resolve this feature into two separate, equally
bright sources (H2 and H4), the more western of which (H2) appears coincident with a
bright stellar-like object, with a B magnitude of 10.7. The apparent optical
counterpart to H4 (not seen in the APM finding charts of Irwin \etal\ (\cite{Irwin})) 
is much fainter (B=14.7, see Fig.\.8(left)). What is rather striking
though, is that, what appears to be a `twin' of H2/4-P4 can be seen on the opposite
side of IC~4329A, at an extremely similar projected distance from the bright central
galaxy. This source, H16-P12, appears coincident with a rather faint (B=17.6)
stellar-like object, less than 2\arcsec\ west of HRI position (Fig.\.8(right)). The
positioning of these two sources with respect to the central bright galaxy, IC~4329A,
is both unusual and intriguing.

On July 6, 1997, during an observing campaign at the 2.2m ESO/MPG telescope at La Silla
observatory, we obtained spectra of these optical candidates using the EFOSC2
spectrograph with grism \#4, a 2\arcsec-wide long slit and the $2048 \times 2048$
$15\mu m$ LORAL CCD. This gave a dispersion of 2 \AA\ per pixel, a spectral coverage
of 4100--7500 \AA, and a spectral resolution of 12 \AA\ FWHM. The seeing was typically
1.5". The data were reduced according to the procedure given in Pietsch \etal\
(\cite{Pietsch88}).

Through these optical observations, we have established that all three sources have
nothing at all to do with IC~4329A, and are in fact Galactic foreground or background objects.
Sources H2 and H4 to the south-west, when compared with Jacoby \etal's (\cite{Jacoby})
library of stellar spectra, appear to be foreground stars of types G3~{\rm V} and
M5~{\rm V} respectively. Furthermore, the north-eastern source looks to be a background
quasar with a redshift of $0.5430 \pm 0.0005$. The HRI count rates measured are
consistent with the X-ray fluxes expected from these source classes.

\begin{figure*}
\unitlength1.0cm 
\vspace{85mm}
\hfill \parbox[b]{18.0cm} 
{\caption{Contours of HRI emission overlayed onto optical digitized sky survey images
for the H2/H4 field (left) and the H16 field (right). The X-ray image has been smoothed
with a Gaussian filter of 10\arcsec\ FWHM, and the contours correspond to 0.0625 and
0.1875\,cts arcsec$^{-2}$.
}}
\end{figure*}

All of the remaining non-bright sources appear to have very close ($<3.5$\arcsec)
optical counterparts, the brightest of which is that associated with H17-P14 (B=14.7).

\section{Results and discussion - the unresolved emission}
\label{sec_disc_unres}

What is very evident from Fig.\,3 is that there appears to be a
great deal of unresolved emission enveloping the sources associated with the two
central galaxies, IC~4329A (P10) and IC~4329 (P8). As seen earlier, this is not seen at
first with the HRI (Fig.\,1), though becomes strikingly clear when
an adaptive filtering technique is applied (Fig.\,2). The fact that
this emission is very obvious in the unfiltered PSPC data, though this fact is not
mentioned in M95, is very encouraging, and to investigate this aspect of the X-ray
emission further, a similar filtering technique was applied to the PSPC data. 
An image (of pixel size 15\arcsec, and over the broad 11-235 range) was formed,
and sections of the image were smoothed using progressively larger Gaussians for lower
intensity pixels. Pixels of amplitude 1 (2,3,4,5,6,7,8) were smoothed with a Gaussian
of FWHM 320\arcsec\ (225,160,115,80,55,40,30)\arcsec. Again, pixels of amplitude greater
than 8 remained unsmoothed, to ensure that the bright point sources were not smoothed
into the background. The resultant image is shown in Fig.\,9. Again,
as in the adaptively smoothed HRI image, a great deal of unresolved emission is
apparent, extending to a much greater radius than is seen in the equivalent HRI image
(9\arcm\ compared to 5\arcm). This is not too surprising, even taking into account the
factor of $\sim2$ deficit in exposure times, as the PSPC is far more sensitive to low
surface brightness diffuse emission than the HRI.

As the PSPC has some spectral resolution, we are able to investigate the spectral
properties of this unresolved emission. What was immediately apparent though,
before any rigorous spectral fitting (as described later) was performed, was that
this unresolved emission appeared markedly two-component. It was decided at first to
investigate the spatial properties of these two components, to try and ascertain 
their distributions.

\subsection{The unresolved emission - spatial properties}

This two component nature is apparent in Fig.\,10. Shown are two images, each
essentially the same as Fig.\,9, but extracted from two separate energy bands, the
first (soft), extracted from channels 8$-$41, the second (hard) extracted from
channels 52$-$201. In both cases, images with 15\arcsec\ resolution were formed and
these images were smoothed using progressively larger Gaussians for lower intensity
regions. Pixels of amplitude 1 (2,3,4,5,6,7,8) were smoothed with a Gaussian of
FWHM 320\arcsec\ (225,160,115,80,55,40,30)\arcsec, and pixels of amplitude greater than
8 were not smoothed, ensuring that the bright point sources were not smoothed into
the background).

The results of this procedure are really rather striking, and the two components,
the soft and the hard components, show remarkably different properties. The soft
component is dominated by IC~4329A and the bright H1-P3 source to the south-west.
Strong features are also seen associated with IC~4329, with the two outlying
sources, H2/4-P4 and H16-P12, and with H15-P11. The {\em unresolved} soft emission
though, which is what we are here most interested in, appears almost entirely to
the south-east of the bright IC~4329A source, in a roughly semi-circular
distribution, `centred' on IC~4329A, and extending out past the two outlying sources
H2/4-P4 and H16-P12. The emission appears far from uniform, containing much
structure, notably in the south-east direction, perpendicular to the IC~4329A disc
(and to the line joining the two outlying sources to IC~4329A). Some evidence
exists for a similar (though very much smaller) extension to the north-west of
IC~4329A, though much of this emission may be due to IC~4329.

\begin{figure}
\unitlength1.0cm 
\vspace{85mm}
\hfill \parbox[b]{8.7cm} 
{\caption{\Ros\ (0.1$-$2.4\,keV) PSPC map of the IC~4329A field obtained using an
adaptive filtering technique (see text), overlayed on a digitized sky survey image. 
The contour levels are at 2, 3, 5, 9, 15, 31, 63, 127, 255,
511, 1023 and 2047$\sigma$ ($\sigma$ being $8.3\times10^{-5}$\,cts s$^{-1}$
arcmin$^{-2}$) above the background ($1.3\times10^{-3}$\,cts s$^{-1}$
arcmin$^{-2}$). 
}}
\end{figure}

As regards the hard emission, many sources, notably IC~4329A, IC4329, H1-P3, and the
two outlying sources, appear as strong sources. The unresolved hard emission however,
appears markedly different to the unresolved soft X-ray structure. Firstly, it extends
not only to the south-east, as the soft component does, but also to the north-west,
and appears roughly circular in nature. Furthermore, in contrast to the soft
component, it appears rather uniform and smooth, with little in the way of
substructure. Interestingly, and as discussed later, the hard component is not centred
on IC~4329A, but instead appears to be centred somewhere between the IC~4329A/IC~4329
pair. Also of note is the fact that this emission is seen to envelope two further
galaxies within the galaxy group mentioned in the introduction. These two galaxies,
IC~4327 to the north-west of the IC~4329A/IC~4329 pair, and NGC~5298 to the south-west,
can be seen at the edge of the hard unresolved emission (Fig.\,10).

Structure in the unresolved emission is evident also in the HRI data. Fig.\,11 shows
contours of HRI X-ray emission superimposed on an optical image of the two central galaxies.
A 3\arcsec\ resolution image was extracted from the HRI channel range 6$-$11, and smoothed
with a Gaussian of FWHM 7\arcsec. Firstly, as regards the above discussion, significant
emission, in what appears to be in the form of a `bridge' (with a curious northern plume),
is seen connecting the two galaxies. Secondly, with regard to the emission surrounding
IC~4329A, there appears to be an elongation along the disc, pointing towards the two
outlying sources, one of which, H2/4-P4, can be seen at the bottom-right of Fig.\,11. Also,
there appears to be good evidence for an extension perpendicular to the IC~4329A disc,
especially, as seen in the PSPC soft band image above, to the south-east. Lastly, as regards
IC~4329, the curious spiral-arm-like structure to the X-ray emission is very intriguing.

\begin{figure*}
\unitlength1.0cm 
\vspace{85mm}
\hfill \parbox[b]{18.0cm} 
{\caption{\Ros\ PSPC maps of the IC~4329A field in the (left) soft (channels 8$-$41)
and (right) hard (channels 52$-$201) bands, obtained using an adaptive filtering
technique (see text), and overlayed on optical images. The contour levels are at 2,
3, 5, 9, 15, 31, 63, 127, 255, 511, 1023 and 2047$\sigma$ ($\sigma$ being
$2.3\times10^{-5}$ (soft) and $4.3\times10^{-5}$ (hard) cts s$^{-1}$ arcmin$^{-2}$)
above the background ($7.1\times10^{-4}$ (soft) and $6.4\times10^{-4}$ (hard) cts
s$^{-1}$ arcmin$^{-2}$).
}}
\end{figure*}

\subsection{The unresolved emission - spectral properties}

It appears therefore, that the unresolved emission within the IC~4329A/IC~4329
system is two-component, and this two component nature varies across the system, the
area to the north-west of the IC~4329A disc being hard, the area to the south-east,
being hard also, but with a very significant soft component contribution. This
north-west/south-east divide was used to investigate the spectral properties of the
unresolved emission. Several unresolved emission spectra were extracted, both from
the north-west (NW) and the south-east (SE) side, a line through the centre of
IC~4329A, at an angle (anticlockwise from north) of 56$^{\circ}$ (so that it
approximately followed both the IC~4329A disc and the line joining the two outlying
sources to IC~4329A), used to separate the two halves. Two main spectra (NW and SE)
were extracted from half-annuli centred on IC 4329A, with inner radii of 2.25\arcm\
and outer radii of 11.7\arcm, and were binned to give a signal-to-noise ratio of
approximately six in each channel. These two spectra were each further subdivided
into three concentrically-extracted spectra, with inner and outer radii of
2.25\arcm$-$5\arcm\ (NW1 and SW1), 5\arcm$-$8.25\arcm\ (NW2 and SW2) and
8.25\arcm$-$11.7\arcm\ (NW3 and SW3). These six sub-spectra were each binned to
give a signal-to-noise ratio of approximately five in each channel. In the
extraction of each of these 8 total spectra, data associated with each of the
sources were excluded to a radius of 1.5\arcm. A background spectrum, free of
unresolved emission, was extracated from a 15\arcm$-$18.25\arcm\ annulus, data
associated with the sources excluded to a radius of 2\arcm. Once these spectra were
corrected for background and exposure time, it was possible to gauge the soft and
hard-band contributions within each area. In the north-west, the percentage of
counts in the soft (0.1$-$0.5\,keV) band, compared to the total (0.1$-$2.4\,keV)
number of counts, is seen to be low ($\approx8.5$\%) and constant (to within
$\sim3$\%) for each of the three spectra (NW1, NW2, NW3). In the south-east however,
the soft band contribution is seen to rise sharply from 19.7\% for the
inner-extracted spectrum (SE1), through 34.8\% for SE2, to 53.6\% for SE3.

As in the fitting of the point source spectra, spectral models were fitted to these
unresolved emission spectra. Dealing first with the north-west spectra, power-law,
and Raymond \& Smith hot plasma models were first attempted, and the results of the
best fits are summarized in Table~\ref{table_fits_nw}. As can be seen, power-law
models (cols.\,2-6) are able to fit not only the NW spectrum, but also the three
separate NW1, NW2 and NW3 spectra, very well, with reduced $\chi^{2}$s of less than
unity (in fact more like 0.5 in every case, apart from NW2). Though the resulting
parameters at first appear rather different, the errors are rather large, and the
spectral fit parameters are quite consistent with one another, \ie the north-western
spectrum appears to remain constant with radius. Fig.\,12 (left) shows the 99\%,
95\% and 68\% confidence contours in the absorbing column$-$spectral index plane
for the power-law fit to the full NW spectrum (\ie\ Table~\ref{table_fits_nw},
row~1).

\begin{figure}
\unitlength1.0cm 
\vspace{85mm}
\hfill \parbox[b]{8.7cm} 
{\caption{\Ros\ HRI map of the IC~4329A field obtained using an adaptive filtering
technique (see text), overlayed on a digitized sky survey image. 
Only channels 6$-$11 are used. 
The contour levels are at 2, 3,
5, 9, 15, 31, 63, 127, 255, 511, 1023 and 2047$\sigma$ ($\sigma$ being
$7.1\times10^{-4}$\,cts s$^{-1}$ arcmin$^{-2}$) above the background
($2.1\times10^{-3}$\,cts s$^{-1}$ arcmin$^{-2}$).
}}
\end{figure}

\begin{table*}
\caption[]{
Results of fitting power-law and Raymond \& Smith hot plasma models to the
north-western unresolved emission spectra (see text), as follows: Column 1 gives the
spectrum, whether the full NW spectrum or one of NW1, NW2, or NW3, the radially
extracted spectra. Columns 2$-$6 give the results of the best power-law fits; fitted
$N_{\rm H}$ (col.\,2), fitted spectral index, $\Gamma$, where $F\propto E^{-\Gamma}$
(col.\,3), and the reduced $\chi^{2}$ and number of degrees of freedom (col.\,4). Columns
7$-$9 give the results of the best Raymond \& Smith hot plasma model fits (with the
metallicity frozen at the solar value); fitted $N_{\rm H}$ (col.\,7), fitted temperature
kT, in keV (col.\,8), and the reduced $\chi^{2}$ and number of degrees of freedom
(col.\,9). Where parameters are seen to `peg' at the highest or lowest allowable values,
2$\sigma$ upper or lower limits are given.
Two values for the (0.1$-$2.4\,keV) X-ray luminosity (scaled to account for
the emission lost in the `holes' left after the source subtraction procedure) are
each given (corresponding to a distance of 64\,Mpc) in columns 5 and 6 (for the
power-law fits) and columns 10 and 11 (for the hot plasma fits). Cols.\,5 and 10 give
the `intrinsic' luminosity (correcting both for Galactic and intrinsic absorption),
while cols.\,6 and 11 give the `emitted' luminosity (\ie correcting merely for the
Galactic $N_{\rm H}$).
}
\label{table_fits_nw}
\begin{tabular}{l|rrrrr|rrrrr}
\hline
\noalign{\smallskip}
Spec. & \multicolumn{5}{c}{Power-law model} & \multicolumn{5}{c}{Raymond \& Smith model} \\ \hline
 & $N_{\rm H}$ & Photon & red.\,$\chi^{2}$ & 
	\multicolumn{2}{c}{$L_{\rm x}$ (10$^{41}$\,erg s$^{-1}$)}   
 & $N_{\rm H}$ & kT       & red.\,$\chi^{2}$ & 
	\multicolumn{2}{c}{$L_{\rm x}$ (10$^{41}$\,erg s$^{-1}$)}   \\
 &(10$^{20}$\,cm$^{-2}$)& Index & (n.d.o.f.) & (Int.) & (Em.) 
 &(10$^{20}$\,cm$^{-2}$)& (keV) & (n.d.o.f.) & (Int.) & (Em.) \\
(1) & (2) & (3) & (4) & (5) & (6) & (7) & (8) & (9) & (10) & (11) \\
\noalign{\smallskip}
\hline
\noalign{\smallskip}
NW & 22.0$^{+36.5}_{-17.5}$ & 2.03$^{+1.65}_{-1.15}$ & 0.55(9) & 26.3 & 8.67
& 10.0$^{+13.5}_{-5.00}$ & $>1.9$ & 0.50(9)    & 8.80 & 6.08 \\
NW1& 3.60$^{+27.5}_{-0.20}$ & 0.96$^{+1.60}_{-0.30}$ & 0.48(5) & 4.69 & 4.69
& 5.90$^{+15.4}_{-3.30}$ & $>1.9$ & 0.69(5)    & 3.75 & 3.24 \\ 
NW2& 48.3$^{+77.0}_{-42.5}$ & 3.15$^{+2.33}_{-2.10}$ & 0.95(7) & 72.8 & 2.90
& 16.4$^{+32.2}_{-10.5}$ & 2.62$^{+11.1}_{-1.00}$ & 0.84(7)    & 3.50 & 2.07 \\
NW3& $<110$ & 1.09$^{+2.15}_{-1.09}$ & 0.41(3) & 3.13 & 3.13
& $<55$ & $>1.6$ & 2.38(3)    & 2.77 & 2.77 \\
NW & 22.7(F)      & 2.06$\pm$0.37& 0.50(10)& 27.4 & 8.63
& 4.40(F)      & $>2.4$ & 0.71(10)   & 7.91 & 7.91 \\
NW1& 22.7(F)      & 1.97$\pm$0.50& 0.75(6) & 11.1 & 3.79
& 4.40(F)      & $>2.2$ & 0.61(6)    & 3.46 & 3.46 \\
NW2& 22.7(F)      & 1.99$\pm$0.51& 0.91(8) & 9.66 & 3.26
& 4.40(F)      & 2.59$^{+7.2}_{-0.6}$ & 1.24(8)    & 2.66 & 2.66 \\
NW3& 22.7(F)      & 2.07$\pm$0.89& 0.62(4) & 7.38 & 2.31
& 4.40(F)      & $>2.0$ & 0.34(4)    & 2.16 & 2.16 \\
\noalign{\smallskip}
\hline
\end{tabular}
\end{table*}

One should note here the bottom four power law fits, given in
Table~\ref{table_fits_nw}, the implications of which are discussed more fully
later. Here, an attempt has been made to see how consistent the NW spectra are with
the spectrum of the bright central source IC~4329A. Upon freezing the absorption
column at the value obtained in the fitting of the IC~4329A spectrum
(22.7$\times10^{20}$\,cm$^{-2}$, see Table~\ref{table_fits_sources}), the quality
of each fit is seen to remain very good. The values obtained for the spectral
index however, though consistent with each other, are somewhat higher than the
IC~4329A value. 

\begin{figure*}
\unitlength1.0cm 
\vspace{85mm}
\hfill \parbox[b]{18cm} 
{\caption{Model spectrum fits to the NW spectrum. (Left) Gaussian contour levels of
1$\sigma$, 2$\sigma$ and 3$\sigma$ in the spectral index$-$absorption column plane for
the power-law fit to the NW spectrum (Table~\ref{table_fits_nw}, col.\,1). (Right)
Gaussian contour levels of 1$\sigma$, 2$\sigma$ and 3$\sigma$ in the
temperature$-$absorption column plane for the Raymond \& Smith plasma fit to the NW
spectrum (Table~\ref{table_fits_nw}, col.\,1).
}}
\end{figure*}

Moving on now to the Raymond \& Smith hot plasma model fits to the NW spectra
(cols.\,~7-11), the quality of the fits is again good, in the majority of cases.
The absorbing column appears to be small and is consistent with the Galactic value
in every case. Unfortunately, as regards the temperature, little definite
information can be gleaned from these results, whether the absorbing column is left
to optimize or is frozen at the Galactic value. Though the fitted temperatures
appear to be high, the fitting procedure often `pegging' the temperature at the
highest allowable value, the errors are large. The size of this error region can be
seen in Fig.\,12 (right), where 99\%, 95\% and 68\% confidence contours are shown
in the absorbing column$-$temperature plane for the Raymond \& Smith hot plasma
model fit to the full NW spectrum (\ie\ Table~\ref{table_fits_nw}, row~1).

Finally, it is worth noting that usage of a thermal bremsstrahlung model in the 
spectral fitting gives very similar results to the Raymond \& Smith results. The 
equivalent absorbing column$-$temperature confidence grid appears essentially 
identical to Fig.\,12 (right). Only in the fitting to the total NW spectrum was 
a best fit realised $-$ $N_{\rm H} = 16.5^{+23.0}_{-11.4}\times10^{20}$\,cm$^{-2}$, kT
$>1.05$\,KeV, 
with a reduced $\chi^{2}$ slightly worse than for the Raymond \&
Smith case (0.54, with 9 degrees of freedom). 

The emission to the south-west, as mentioned previously, is considerably more complicated,
requiring a two-component model to fit the spectra, even adequately (the best one-component
power-law and Raymond \& Smith hot plasma fits to the SE spectrum result in reduced $\chi^{2}$s
of over 5 and 7, respectively). Freezing of a number of the components was necessary, as too
many free parameters led to fits that either refused to settle at consistent values or had huge
error regions. Following the line of thought touched upon above, the hard component was assumed
to be identical to the IC4329A spectrum. The variation of the spectral parameters can be seen
in Fig.\,13. Here, the 99\%, 95\% and 68\% confidence contours in the soft-component absorption
column$-$temperature plane are shown for the full SE spectrum, \ie the hard
component is frozen to that of IC4329A, and the metallicity of the soft component 
is frozen at the solar value.

\begin{figure}
\unitlength1.0cm 
\vspace{85mm}
\hfill \parbox[b]{8.7cm} 
{\caption{Gaussian contour levels of
1$\sigma$, 2$\sigma$ and 3$\sigma$ in the soft-component temperature$-$absorption
column plane for the (hard) power-law plus (soft) Raymond \& Smith plasma fit to the
SE spectrum. }}
\end{figure}

As can be seen, the soft component appears to be very soft, and rather unabsorbed. 
Fitting of the three sub-spectra SE1, SE2 and SE3, failed to settle at consistent
values, and it is only when the absorption column of the soft component is fixed at 
the Galactic value (4.40$\times10^{20}$\,cm$^{-2}$), as seems completely reasonable 
from Fig.\,13, that good fits are obtained. 
Table~\ref{table_fits_se} summarizes these best two-component (power-law plus Raymond
\& Smith hot plasma) model fits to the SE spectra. 
In actuality, whatever (within reason) the hard component is frozen at (for
instance, the higher index fit to the NW spectra - see Table~\ref{table_fits_nw}),
or whether the soft-component metallicity is frozen or left free, makes very little
difference to the resultant fits. Assuming therefore that a hard component somewhat
similar (if not identical) to the IC4329A spectrum is present, then a soft
component, consistent with a very cool, relatively (if not completely) unabsorbed
hot plasma, appears to be present. It is also worth noting here, that the equivalent
confidence contour plots for the three sub-spectra SE1, SE2 and SE3, appear
more-or-less identical, though with larger-spaced contours reflecting the reduction
in statistics.

Models incorporating two Raymond \& Smith hot plasmas either failed to converge or, if they
did, the results obtained were unphysical or had uncomfortably large error regions. It is worth
noting that a model incorporating two thermal bremsstrahlung models, both with absorbing
columns frozen at the Galactic value, is able to fit the SE spectrum. A very low and a very
high temperature plasma is required.

\begin{table*} 
\caption[]{
Results of fitting two-component (power-law + Raymond \& Smith hot plasma) models to
the south-eastern unresolved emission spectra (see text), as follows: Column 1 gives
the spectrum, whether the full SE spectrum or one of SE1, SE2, or SE3, the radially
extracted spectra. Columns 2 and 3 give the parameters of the power-law component to
the best fit; fitted $N_{\rm H}$ (col.\,2), and fitted spectral index, $\Gamma$, where
$F\propto E^{-\Gamma}$ (col.\,3). Columns 4 and 5 give the parameters of the Raymond
\& Smith hot plasma component of the best fit (with the metallicity frozen at the
solar value); fitted $N_{\rm H}$ (col.\,4), and fitted temperature kT, in keV (col.\,5).
The reduced $\chi^{2}$ and number of degrees of freedom are given in (col.\,6). Two
values for the (0.1$-$2.4\,keV) X-ray luminosity (scaled to account for the emission
lost in the `holes' left after the source subtraction procedure) are given
(corresponding to a distance of 64\,Mpc) in columns 7 and 8. col.\,7 gives the
`intrinsic' luminosity (correcting both for Galactic and intrinsic absorption), and
col.\,8 gives the `emitted' luminosity (\ie correcting merely for the Galactic $N_{\rm
H}$), the figures in brackets indicating the percentage contribution to this
luminosity from the Raymond \& Smith hot plasma (\ie the soft) component.
}
\label{table_fits_se}
\begin{tabular}{lrrrrrrr}
\hline
\noalign{\smallskip}
Spectrum & \multicolumn{2}{c}{Power-law component} & \multicolumn{2}{c}{Raymond \& 
Smith component} & red.\,$\chi^{2}$ & 
\multicolumn{2}{c}{$L_{\rm x}$ (10$^{41}$\,erg s$^{-1}$)}   \\ 
 & $N_{\rm H}$          & Photon & $N_{\rm H}$           & kT   &  (n.d.o.f.) &
(Int.) & (Em.) \\
 &(10$^{20}$\,cm$^{-2}$)& Index    & (10$^{20}$\,cm$^{-2}$)& (keV)&             \\
(1) & (2) & (3) & (4) & (5) & (6) & (7) & (8) \\
\noalign{\smallskip}
\hline
\noalign{\smallskip}
SE & 22.7(F)& 1.28(F)& 4.40(F)& 0.11$\pm$0.05 &1.57(7) & 17.2$\pm$1.0 (40\%) 
	& 12.5$\pm$0.8 (54\%) \\
SE1& 22.7(F)& 1.28(F)& 4.40(F)& 0.11(F)       &1.64(5) & 6.05$\pm$0.6 (18\%) 
	& 3.83$\pm$0.3 (28\%) \\
SE2& 22.7(F)& 1.28(F)& 4.40(F)& 0.13$\pm$0.04 &0.87(6) & 6.87$\pm$0.6 (37\%) 
	& 4.95$\pm$0.4 (52\%) \\
SE3& 22.7(F)& 1.28(F)& 4.40(F)& 0.11$\pm$0.07 &1.19(3) & 4.48$\pm$0.7 (62\%) 
	& 3.73$\pm$0.6 (75\%) \\
\noalign{\smallskip}
\hline
\end{tabular}
\end{table*}

As we have seen therefore, both a hard and a soft component to the unresolved
emission appear to exist around IC~4329A, the hard component lying, rather smoothly
and symmetrically distributed, around the IC~4329A/IC~4329 pair, the soft component
lying almost entirely to the south-east of IC~4329A. As we have regions of emission
where only the hard emission appears to exist (to the north-west), it is easier to
deal with this aspect of the emission first. Only after this, can we move
on to discuss the soft emission component.

\subsection{Discussion - The hard component}

One possibility as to the origin of the hard residual component, as alluded to
(though not explicitly stated) in the previous sections, is that it could be due to
the `wing' emission from the very bright central source. This idea is supported by
the fact that the spectrum of the hard residual emission (both the NW emission and
the hard component of the SE emission) appears very consistent with the IC~4329A
spectrum. 

To investigate this question further, a radial surface brightness profile (over the
channel range 11 to 235) of the central emission was formed. This is shown in
Fig.\,14 (with the region between 2.4\arcm\ and 12.6\arcm\ magnified in the
inset), and contains many features, as follows: Firstly, the data points show the
radial distribution (centered on IC~4329A) of the total X-ray emission, split into
21\arcsec\ radial bins, with the region inside 1\arcmin\ split into 7\arcsec\
radial bins. The long dashed line represents the PSF of a point source
with a spectrum best fitted by a power law of photon index 1.28, absorbed by a
column of 2.27$\times10^{21}$\,cm$^{-2}$, \ie it represents  the radial
distribution of the emission from IC~4329A, assuming it to be a point source. This
was formed by summing together model PSFs (for photons of varying energies), each
normalized according to the spectrum of IC~4329A (\ie normalized according to
Fig.\,7). The dotted line indicates the radial distribution of the true (\ie
non-diffuse) background. This was formed by firstly eliminating from an image all
of the sources to very large radii (5$\times$ the FWHM of the PSF). This resulted
in an image where approximately none of the unresolved, extended emission, visible
in Fig.\,9, was included. A polynomial fit was then used to interpolate across the
`holes', and the image was then heavily smoothed (with a Gaussian of FWHM
9.4\arcm). A radial profile of this image (again centered on IC~4329A) gives the
indicated dotted line. The dash-dotted line indicates the level of the `diffuse'
emission, the unresolved emission visible in Fig.\,9. This was formed by removing
sources from an image to radii of twice the FWHM of the PSF. A mask image was
created in the same way with zero values at the positions of these holes, and
values of unity everywhere else. A radial profile (of binsize 36\arcsec) was formed
from the source subtracted image, and this was normalized for the area lost in the
`holes', by dividing it by an equivalent profile of the mask image.

\begin{figure*}
\unitlength1.0cm 
\vspace{125mm}
\hfill \parbox[b]{18.0cm} 
{\caption{Surface brightness profile of the (0.1$-$2.4\,keV) flux about the centre of
IC~4329A (the region between 2.4\arcm\ and 12.6\arcm\ is magnified in the inset). 
Data points show the total X-ray emission profile, the long dashed line 
represents the PSF of a point source with spectral properties identical to that of 
IC~4329A, the dash-dotted line indicates the level of the `diffuse' emission, all 
point source emission having been removed, and the dashed line indicates the level of 
the `true' background, all point source and `diffuse' emission having been removed (see
text).
}}
\end{figure*}

Several features are evident within this figure. The central bright point source is
very obvious, but also visible are the two other bright sources, the IC~4329 source
(H5-P6), at $\approx$3\arcm, and H1-P3, at $\approx$13.5\arcm. The emission is seen
to fall to the background level at large radii (9$-$18\arcm), but there is a  very
notable deviation from this level closer in. Between 4\arcm\ and 9\arcm\ (and
possibly up to 11\arcm), the emission is seen to be significantly enhanced with
respect to the background, and there appears to be two components to this
enhancement. One is the presence of some underlying, unresolved, perhaps truly
diffuse emission, as indicated by the dash-dotted line. The second is the presence
of point sources, notably the bright sources P6, P9 and P16, visible in the radial
profile figure, between 4.5\arcm\ and 6.5\arcm, as data points lying above the
residual emission profile. Part (or perhaps all) of the enhancement at
$\approx10$\arcm\ is due to sources P3 and P4.

With regard to the unresolved emission therefore, it is the dash-dotted, unresolved
emission profile, and the possible contribution to this from the dashed IC~4329A point
source that are of interest. It can be seen from Fig.\,14 that the idealised point
source model of IC~4329A does contribute to the unresolved emission within the
inner annulus (NW1 \& SE1: 2.25\arcm$-$5\arcm), though beyond this, in the outlying
regions, the expected contribution drops rather quickly. The fact that the X-ray source
associated with IC~4329A may not be an idealised point source however,
as is suggested by Fig.\,14, where some deviation from the (spectrally
corrected) point source PSF model can be seen), may boost the true wing contribution in
the outlying regions from this source, to values perhaps in line with that of the
residual emission.

Perhaps therefore, the hard residual emission {\em is} due to the wings of IC~4329A.
The hard residual emission is well fit by the IC~4329A spectrum, and the number
of counts expected from the IC~4329A wings may be close to being in agreement with
what is observed (though this is unlikely). 
Furthermore, the hard residual emission appears approximately
circular, and both smooth and regular, as one would expect, if this emission were
just due to the wings of a bright point source. Unfortunately, there is one aspect
of the hard residual X-ray emission that is very difficult to reconcile with it
being due to the IC~4329A wings $-$ the emission is {\em not} centred on IC~4329A.
Instead, it appears centred on a point midway between IC~4329A and its bright
elliptical companion, IC~4329, a point coincident with the `bridge' of X-ray
emission visible in the HRI data (Fig.\,11). This fact, that the emission is not
centred on IC~4329A, makes it very unlikely that the hard residual emission is due
entirely to the IC~4329A wings.

Instead, the above points are very suggestive of the emission being due 
partly to the `wings' of IC~4329A, and partly to
hot gas surrounding the IC~4329A/IC~4329 pair. A significant fraction of the
hard residual emission will be due to the IC~4329A wings, but this only makes a
marked contribution at smaller radii, dropping to a level below half that of the
total hard residual emission beyond $\approx4$\arcm. The fact that the spectral
properties of the hard residual emission appear indistinguishable from those of
IC~4329A may at first seem strangely coincidental. In fact, this is not too
surprising as the spectral capabilities of the \Ros\ PSPC are not too good for
harder spectra, and, to the PSPC, an IC~4329A-type spectrum and a hot (5$-$10\,keV)
plasma spectrum appear almost identical, as is apparent in the spectral fitting
results (Table~\ref{table_fits_nw}).

Nevertheless, we have attempted to further analyse the north-western unresolved
emission spectra, in an attempt to extract any 
information regarding the hot gas component that may exist. A two-component model,
comprising of a power-law component, representing the emission from the IC~4329A 
`wings', and a Raymond and Smith hot plasma component, representing the diffuse
hot gas component, was fitted to the full NW spectrum and to the three sub-spectra 
NW1, NW2 and NW3. Here, the power-law component parameters were frozen at the 
IC~4329A values ($N_{H} = 22.7\times10^{20}$\,cm$^{-2}$, $\Gamma=1.28$), and the 
Raymond \& Smith component absorption column was frozen at the Galactic value 
($4.4\times10^{20}$\,cm$^{-2}$). Examination of the resultant Raymond \& Smith
component normalization-temperature parameter space reveals a similar effect in
all but one of the cases; In the fits to the full NW spectrum, and to the inner NW1
and the outer NW3 spectrum, the temperature of the hot gas component 
is seen to be very ill-defined, with large, and highly irregularly-shaped
confidence contour levels. Furthermore,
no significant improvements in the fit quality are seen, reduced $\chi^{2}$s (and
number of degrees of freedom) for these three fits being: NW 0.86(9), NW1 0.63(5)
NW3 0.30(3) (compare these values with the one-component Raymond and Smith values
given in Table~\ref{table_fits_nw}).

In the case of the NW2 spectrum however, a significant improvement is seen in the
fit quality (reduced $\chi^{2}$(n.d.o.f) $= 0.82(7)$ $-$ compare with 
Table~\ref{table_fits_nw}). Also the hot gas component normalization-temperature
parameter space is seen to be quite well-defined, with (at least at the 68\% 
confidence level) reasonably small and regularly-shaped contour levels (see
Fig.\,15). The best fit to the NW2 spectrum (indicated by the dot in Fig.\,15) 
contains a IC~4329A-like component and a hot gas component with a temperature of
1.53$^{+4.35}_{-0.55}$\,keV. 

We believe that we are only able to see this hot gas component, even reasonably,
within the NW2 fit, because it is only at this radial distance from the central
bright source that the contamination from the IC~4329A `wings' drops sufficiently
enough to allow the hot-gas component (which is itself dropping with radius) to
become visible. In the NW1 spectrum, although the hot gas component is large, the
IC~4329A wing component is huge, and the contamination is too large to allow a
reasonable determination of the hot gas component parameters. In the NW3 spectrum,
even though the IC~4329A wing component has become extremely low, the hot gas
component is now very low itself, and the statistics involved are insufficient for
a reasonable spectral determination.

A great range in temperature is able to fit the hot gas component of the full NW
spectrum. We assume here though, that the temperature of the entire hot gas
component is equal to the temperature found for the NW2 spectrum
(1.53$^{+4.35}_{-0.55}$\,keV). 
Fitting of the full NW spectrum, assuming this temperature, does lead to a very
good fit indeed, with a reduced $\chi^{2}$ of 0.52 (with 10 degrees of freedom).

\begin{figure}
\unitlength1.0cm 
\vspace{85mm}
\hfill \parbox[b]{8.7cm} 
{\caption{Two-component model spectrum fit to the NW2 (middle) spectrum. Gaussian
contour levels of 1$\sigma$, 2$\sigma$ and 3$\sigma$ in the hot gas (Raymond \&
Smith) component temperature$-$normalization plane for the (IC~4329A-like) 
power-law plus Raymond \& Smith plasma fit to the NW2 spectrum (see text).
}}
\end{figure}

Mean physical properties for this north-western hot gas can be inferred from 
the above results if we make some assumptions about the geometry of the emission.
Here we have assumed the simple geometry of the north-western hot gaseous emission
being hemispherical with a radius of 11.7\arcm\ (in actuality, only a rough
approximation to the gas properties can be calculated here and assumption of a
slightly different radius gives rise to very similar results).

Using the volume derived for this hemispherical `bubble' model, the fitted emission
measure $\eta n_{e}^{2} V$ (where $\eta$ is the `filling factor' -- the fraction of
the total volume $V$ which is occupied by the emitting gas) can be used to infer
the mean electron density, $n_{e}$, and hence the total mass $M_{\mbox{\small
gas}}$, thermal energy $E_{\mbox{\small th}}$ and cooling time $t_{\mbox{\small
cool}}$ of the gas. 

Performing these calculations, after first accounting for the extra emission lost
in the `holes' left after the source-subtraction procedure, one arrives at
approximate values to the physical properties of the north-western hot gaseous
emission as follows; 
X-ray luminosity $L_{X}$ (0.1$-$2.4\,keV; intrinsic) 2.6$\times10^{41}$\,erg s$^{-1}$;
$n_{e}$, 1.8$\times10^{-4} \eta^{-0.5}$\,cm$^{-3}$;
$M_{\mbox{\small gas}}$, $9.5\times10^{10} \eta^{0.5}$\, $M_{\odot}$;
$E_{\mbox{\small th}}$, 8.1$\times10^{59} \eta^{0.5}$\,erg; 
$t_{\mbox{\small cool}}$, 99\,Gyr. 
In the later discussion, we will see that no spectral
information regarding the hot gaseous emission to the south-east can be obtained.
This is primarily because of this emission being severely contaminated, not only,
as it is to the north-west, by the IC~4329A `wings', but also by the very soft 
emission discussed earlier. If one makes the simple assumption that the properties
of the hot gas to the south-east are identical to those to the north-west, a quite
valid assumption, given the physical appearance of the emission and the level of
accuracy we can attain, then the values of the physical properties for the total
hot gas surrounding the IC~4329A/IC~4329 system become double those quoted above
(except for the mean electron density $n_{e}$, and the cooling time
$t_{\mbox{\small cool}}$, which remain unchanged).

Remembering that, as discussed in the introduction, the two central galaxies lie
close to cluster A3574, and are part of a loosish group of seven galaxies, 
it is useful to compare here the properties of this hot gaseous emission 
component with the general X-ray properties of groups and clusters. 

As regards the comparison with clusters, though the temperature of the hot gas may
be comparable, the luminosity ($5.2\times10^{41}$\,erg s$^{-1}$) is extremely low.
Clusters typically have X-ray luminosities greater than $5\times10^{43}$\,erg
s$^{-1}$ (Edge \& Stewart \cite{Edge}; Yamashita \cite{Yamashita}; White 
\cite{White}), \ie two orders of
magnitude greater than is observed around IC~4329A/IC~4329, and it is therefore 
very unlikely that we are seeing emission associated with the A3574 cluster. In
the following discussion, we will primarily deal with the comparison of the present
results with the properties of galaxy groups.

The general X-ray properties of galaxy groups have recently been published, both of
Hickson's (\cite{Hickson82}) compact groups (HCGs) (Ponman \etal\ \cite{Ponman}), and of other, poor
groups (Mulchaey \etal\ \cite{Mulchaey95}). 
The group containing IC~4329A and
IC~4329 appears rather loose and contains perhaps seven members (a rather large
number). A good optical image of the group can be found in Fricke \& Kollatschny
(\cite{Fricke}). 

The X-ray luminosity of the hot gaseous emission is low, but in no way
uncomfortably low when compared to the results of Ponman \etal\ (\cite{Ponman}) and
Mulchaey \etal\ (\cite{Mulchaey95}). Many other galaxy groups at the same
distance are seen with very similar hot gas X-ray luminosities (\ie the emission
from the member galaxies having been removed, as has been done here). Furthermore,
the ratio of the diffuse hot gas X-ray luminosity to the total blue luminosity of
the member galaxies $L_{X}/L_{B}$ ($\approx1.2\times10^{-3}$) is seen to be very
typical of groups. As regards how the X-ray luminosity relates to the spiral
fraction (the fraction of the group member galaxies that are spiral, as opposed to
elliptical galaxies), the hot gas around IC~4329A/IC4329 is again, in no way
unusual. The spiral fraction of the IC~4329A group is actually very high, IC~4329
being the only non-spiral member, and though Ponman \etal\ (\cite{Ponman}) find only a very
weak correlation between spiral fraction and $L_{X}$, Mulchaey \etal\ (\cite{Mulchaey95})
find a somewhat stronger correlation, all of the
groups they detect having an extended X-ray emitting intragroup medium having a
high percentage of early-type galaxies. The emission within the IC~4329A group (a
high spiral fraction group) is of a low luminosity, in agreement with the work
of Mulchaey \etal\ (\cite{Mulchaey95}) (and of Ponman \etal\ (\cite{Ponman}) though only a far weaker
correlation is seen here).

As regards the temperature of this hot gas however the emission may be 
unusual. As discussed earlier, it has proved very difficult to separate
this emission from the wings of the bright central source associated with
IC~4329A, and a good deal of caution should be taken here.
Indeed, the present case is very similar to the case of HGC~4, as, while
Saracco \& Ciliegi (\cite{Saracco}) suggest the existence of an extended component, Pildis
\etal\ (\cite{Pildis}) find that it is impossible to detect any extended diffuse
component on account of the emission from the central active galaxy in the group
being so strong. 

Our tentative value though, for the temperature of the hot gas ($\approx1.5$\,keV)
appears to be high when compared to the values of Ponman \etal\ (\cite{Ponman}), which range,
in the majority of cases, from 0.6$-$1\,keV, and also when compared to the values of
Mulchaey \etal\ (\cite{Mulchaey95}). Though a large error on the temperature does exist (see
Fig.\,15), it does appear that the temperature is constrained to be greater than
1\,keV. The hot gaseous emission seen surrounding the IC~4329A/IC~4329 pair appears
therefore, to be quite hot when compared to typical groups, and as such, it sits
rather uncomfortably on the $L_{X}-$temperature relationship of Ponman \etal\
(\cite{Ponman}), being of too low a luminosity for its temperature (or too hot for its
$L_{X}$).

For rich clusters of galaxies (Edge \& Stewart \cite{Edge}), there appears to be a good
correlation between $L_{X}$ and the optical velocity dispersion $\sigma$.
Extending the Edge \& Stewart (\cite{Edge}) cluster relationship down, one would expect,
in the case of groups, the higher velocity dispersion systems to have higher X-ray
luminosities, and indeed, this does appear to be the case (Ponman \etal\ \cite{Ponman};
Mulchaey \etal\ \cite{Mulchaey95}); higher-$\sigma$ groups are brighter in X-rays. 
The IC~4329A group of galaxies has a rather large
value of $\sigma$, approximately 390\,km s$^{-1}$, calculated using the
Galaxy-corrected recession velocities given in the Third Reference Catalogue of 
Bright Galaxies (RC3; de Vaucouleurs \etal\ \cite{RC3}). This value is greater (though
within the range of) Hickson \etal's (\cite{Hickson92}) study of 100 HCGs, which have a median
value of 200\,km s$^{-1}$. It is also larger than typical values for loose groups
(208\,km s$^{-1}$ Geller \& Huchra \cite{Geller}; 183\,km s$^{-1}$ Maia \etal\ 
\cite{Maia}), though much less than for rich clusters, where, for example, Zabludoff 
\etal\ (\cite{Zabludoff}) established a median $\sigma$ for 65 Abell clusters, of 744\,km
s$^{-1}$. Interestingly, there is good indication of a correlation between $\sigma$ and
the group emission temperature (Ponman \etal\ \cite{Ponman}), a correlation which appears to
extend from the poorest of groups to the richest of clusters. IC~4329A, on account of
its high temperature {\em and} its high velocity dispersion, lies directly on Ponman
\etal's (\cite{Ponman}) $\sigma-T$ regression line, between the groups and the clusters.
The hot gaseous emission within the IC~4329A group however, 
is not as X-ray luminous as the group's velocity dispersion would imply, and 
it lies below the regression line fitted to the $L_{X}-\sigma$
data of Ponman \etal\ (\cite{Ponman}). It should be stressed however, that it does lie well
within the scatter of the data, and cannot be called in any way, unusual. 

In terms of the amount of gas present, the IC~4329A group appears again, quite 
normal. Its estimated gas mass of $\sim2\times10^{11} M_{\odot}$ is very typical
of groups (Mulchaey \etal\ \cite{Mulchaey95}), lying on the low side of average, but well 
within the scatter. 

In summary then, part of the emission detected surrounding IC~4329A and IC~4329, and
extending out to two other galaxies in the group, IC~4327 and NGC~5298, may well be hot
gaseous emission associated with the galaxy group. If it is, then the estimated
temperature of the hot gas ($\gtsim1$\,keV, quite hot for groups) agrees well with the
(high) velocity dispersion of the group galaxies. Its X-ray luminosity however, appears
to be lower than average (for its velocity dispersion and/or temperature). Similarly, the
mass of gas involved is perhaps lower than average. In no way though, do any of the
physical properties of the emission indicate that this IC~4329A group emission is very
unusual.

Interestingly, the single galaxy group in Ponman \etal's (\cite{Ponman}) survey most
similar to the IC~4329A group is HCG~48. This group has a high velocity dispersion as
well, and a correspondingly high temperature ($\approx1.1$\,keV) is found for the hot
gas. It's X-ray luminosity however, is low, and it is seen to lie significantly below
the $L_{X}$-temperature relationship of Ponman \etal\ (\cite{Ponman}), as do the
present IC~4329A group results. Importantly, like the IC~4329A group, HCG~48 itself
lies within a larger cluster, A1060, and it appears to be falling into this cluster. It
may well be the case that we are seeing in both cases, the HCG~48 case and the IC~4329A
case, the effects of gas stripping from these groups, as they fall through their
respective clusters centres. This reduction in hot group gas would have the effect of
reducing the observed $L_{X}$ (and estimated gas mass), as is observed.

\subsection{Discussion - The soft component}
	
The soft residual component reveals itself in several different ways. It can be seen
in the soft band (Ch.~8$-$41) image of Fig.\,3, where the emission to the
south-east of the central bright IC~4329A source is seen to be significantly
enhanced with respect to the emission to the north-west. The most striking depiction
of the residual soft component however, is in Fig.\,10, where the adaptively
filtered soft band image shows what appears to be a roughly semi-circular
distribution of diffuse emission, `centred' on IC~4329A, and extending out, though
almost entirely to the south-east, to a radius of perhaps 10\arcm, past the two
outlying sources H2/4-P4 and H16-P12. As seen, the spectral properties of this
emission indicate that it is very likely to be due to a very low temperature
($\sim0.1$\,keV) plasma, absorbed by a column consistent with that 
out of own Galaxy.

What is this soft emission then? One question that needs to be addressed first is,
is this feature actually associated with IC~4329A? Perhaps it is a foreground
feature. Perhaps it is very local, within our own Galaxy, or even within our local
bubble. The spectral information present in the PSPC data is very useful here, as
in Fig.\,13, where the 1$\sigma$, 2$\sigma$ and 3$\sigma$ Gaussian contour
levels in the soft-component temperature-absorption column parameter space are
shown, it can be seen that this component's spectrum appears to be best fit with a
plasma spectrum with an absorption column equal to, or greater than, the column
out of the Galaxy, \ie the feature is very likely to be extragalactic, if no
intrinsic absorption is present, a likely fact, given the amorphous, diffuse
nature of the feature. The feature therefore, appears likely to be extragalactic. Is
it associated with IC~4329A though? A couple of points regarding the structure of
the feature are very suggestive, we believe, of this being the case. Firstly, a
good correlation is seen between this soft feature and the 
disc of IC~4329A. Secondly, within this
feature, a large plume of emission is visible lying directly perpendicular to the
IC~4329A disc, to the south-east (see Fig.\,10), \ie along the galaxy's minor axis.
This plume is in the same direction as the 2.3\,GHz radio feature discovered by
Blank \& Norris (\cite{Blank}). Furthermore, the HRI image (Fig.\,11) shows good evidence
for an extension in this south-eastern direction also. Lastly, the soft feature,
although lying almost exclusively to the south-east of IC~4329A, is very symmetric
with respect to the minor axis of IC~4329A, \ie the level of emission seen to the
north-east of the plume discussed above appears very similar to that to the
south-west.

Assuming therefore, that the emission is connected in some way to the IC~4329A/IC~4329
system, then the fact that it is soft and intrinsically, almost completely unabsorbed,
is very reminiscent of the galactic winds seen within the halos of many different
systems (notably starburst systems) (Watson \etal\ \cite{Watson}; Fabbiano 
\cite{Fabbiano88};
Pietsch \cite{Pietsch92}; Strickland \etal\ \cite{Strickland}; Read \etal\ 
\cite{Read97PS}). Indeed, many
multi-wavelength studies of this system indicate that a wind is possibly present (see
Sect.\,\ref{sec_intro}). Both Colbert \etal\ (\cite{Colbert}) and Mulchaey \etal\ 
(\cite{Mulchaey96}) detect $H{\alpha} +$[\nii] features extending along the minor axis, to
$\sim$10\arcsec\ (3\,kpc) on both sides of the nucleus, and both sets of authors believe
that these features represent an outflow from the nucleus similar to the superwinds
commonly seen in edge-on infrared-luminous galaxies (\eg McCarthey \etal\ 
\cite{McCarthey}; Armus \etal\ \cite{Armus}).

There are however, many differences between this feature and the `classic'
winds seen in such nearby starburst systems as M82 and NGC~253. Firstly, the feature
is very large, extending to a radius of perhaps 10\arcm, which at the assumed distance
of 64\,Mpc, corresponds to a size of almost 200\,kpc. This is over an order of
magnitude greater than the classic features mentioned above. Secondly, it appears
almost entirely on one side only of the system, to the south-east, and thirdly, 
it is rather luminous when compared to the classic starburst winds. 

As for the hot gaseous emission in the previous section, mean physical properties
for this cool, soft-component SE gas can be inferred, if we make some assumptions
about the geometry of the emission. We could consider the emission to be in the
form of a hemisphere, purely in the south-eastern direction. This is physically
unreasonable however, and upon closer inspection, the best simple estimate for the
geometry of the soft residual emission appears to be a sphere, centred to the
south of IC~4329A, with a radius of $\approx$8\arcm (150\,kpc at the assumed
distance of 64\,Mpc). Again, only a rough approximation to the gas properties can
be calculated here and assumption of either model gives rise to very similar
results.

Using the volume derived for this spherical `bubble' model, the fitted emission
measure $\eta n_{e}^{2} V$ can again be used to infer the mean electron density,
$n_{e}$, the total mass $M_{\mbox{\small gas}}$, the thermal energy
$E_{\mbox{\small th}}$ and the cooling time $t_{\mbox{\small cool}}$ of the gas,
as follows; 
X-ray luminosity $L_{X}$ (0.1$-$2.4\,keV; intrinsic) 8.7$\times10^{41}$\,erg s$^{-1}$;
$n_{e}$, 3.4$\times10^{-4} \eta^{-0.5}$\,cm$^{-3}$; 
$M_{\mbox{\small gas}}$, $1.1\times10^{11} \eta^{0.5}$\, $M_{\odot}$; 
$E_{\mbox{\small th}}$, 7.0$\times10^{58} \eta^{0.5}$\,erg;
$t_{\mbox{\small cool}}$, 2.6\,Gyr. 

Comparison of these values with values obtained for the soft extended features seen in
several nearby spiral systems (including the `classic' winds in NGC~253, M82, NGC~3079
and NGC~3628 (Read \etal\ \cite{Read97PS})), leads to the conclusion that the soft emission seen
here, to the south-east of NGC~4329A, is even more unusual than at first thought. Not
only, as mentioned above, is it much larger and brighter (both by at least an order of
magnitude) and observed on only one side of the system, but it appears to be far less 
dense (again by over an order of magnitude), and extremely massive (perhaps a hundred
times moreso than for the NGC~253/M82 features). One must be careful here though, as a 
factor of $\sqrt{\eta}$ is still involved in the above results, and this could have 
quite a large effect on the results. 

The soft feature seen here therefore, appears to be completely unlike the classic starburst winds
(except in terms of the temperature, which appears to be consistent with the classic wind
temperatures (Read \etal\ \cite{Read97PS}). What this feature bears much more of a resemblance to,
are the features seen associated with the ultraluminous far-infrared galaxies (FIRGs) Arp~220,
NGC~2623 (Read \& Ponman \cite{Read97P}) and NGC~6240 (Fricke \& Papaderos \cite{Fricke}). Here,
large, soft, outlying features are seen to one side, and one side only, of these high-luminosity
systems, systems believed to be at the most energetic stage in the evolution of a merger between
two galaxies, \ie at the point where the nuclei of the two galaxies merge. However, the features
seen in these systems, though more similar to the IC~4329A feature than the `classic' wind
features, are, in some respects, still rather different. Though they are spectrally soft
($kT\ltsim0.5$\,keV), and large ($\ltsim50$\,kpc), much larger than the `classic' wind features,
they are still significantly smaller than the IC~4329A feature. Furthermore, though they are far
brighter than the `classic' wind features, they still have luminosities perhaps an order of
magnitude smaller than the IC~4329A feature. It must be borne in mind here though that, as
mentioned previously, IC~4329A is a superluminous Seyfert galaxy, with an X-ray luminosity some
two orders of magnitude greater than that of Arp~220 (Read \& Ponman \cite{Read97P}). In this
respect, the IC~4329A `wind' feature makes up a far smaller fraction of the total X-ray luminosity
than in any of the FIRGs or starbursts. 

The most striking similarity between the IC~4329A soft feature, and the FIRG
soft features however, is perhaps the strangest facet of their emission $-$ their one-sided
nature. Why does the structure seen in IC~4329A appear only on one side of the system?
As regards the FIRGs, and as is discussed in Read \& Ponman (\cite{Read97P}), the answer may lie
in the fact that these systems are rapidly evolving. MacLow \& McCray (\cite{MacLow})
numerically modelled the growth of superbubbles: large thin shells of cold gas
surrounding a hot pressurized interior, - essentially a galactic wind, or at least,
the progenitor of, in various stratified atmospheres. They discovered that
superbubbles blow out of the \hi\ layer (\ie they depart from the `snowplow' phase 
and evolve into the `blowout' phase), and are then able to move out into the
inter-galactic medium at velocities of several thousand km s$^{-1}$ (see Heckman
\etal\ \cite{Heckman}). This, MacLow \& McCray (\cite{MacLow}) find, occurs when the superbubbles have a
radius of between one and two scaleheights. What they also find however, is that these
bubbles will blow out on {\em one side only} of a disc galaxy if the bubble centres are
more than 50$-$60\,pc from the centre of the disc of the galaxy. 
Now, in a (relatively) non-evolving system, such as M82 or NGC~253, the starburst is seen to be
very symmetrically positioned with respect to the galactic disc, and bipolar structures are seen
in the X-ray (Watson \etal\ \cite{Watson}; Fabbiano \cite{Fabbiano88}; Pietsch
\cite{Pietsch92}; Strickland \etal\ \cite{Strickland}; Read \etal\ \cite{Read97PS}).
As discussed in Read \& Ponman (\cite{Read97P}) though, in the rapidly-evolving ultraluminous
merging systems Arp220 and NGC~2623, the central burst is highly unlikely to be so centrally
positioned with respect to the quickly-moving and highly distorted gaseous components, and the
direction of steepest pressure gradient (along which the bubble will most rapidly expand) will now
no longer be along both directions of the disc's minor axis (as is the case in M82-type systems),
but will be in just one direction (as is predicted by MacLow \& McCray \cite{MacLow}), hence the
observed one-sided blowout.

In the present case, it appears unambiguous that some sort of interaction between IC~4329A and
IC~4329 is taking place: Both galaxies are close together  and are part of a looser group of seven
galaxies. IC~4329 appears to be a shell galaxy, a signature of interaction. Photographic
co-addition and contrast enhancement of four UKST IIIa-J survey plates indicate the presence of
low surface brightness features around IC~4329A (Wolstencroft \etal\ \cite{Wolstencroft}), again
suggestive of an interaction. Evidence for possible interaction-induced activity is also seen, as
discussed earlier, in the $H{\alpha} +$[\nii] studies of Colbert \etal\ (\cite{Colbert}) and
Mulchaey \etal\ (\cite{Mulchaey96}), and in the radio studies of Unger \etal\ (\cite{Unger}) and
Blank \& Norris (\cite{Blank}). 
Good evidence for an interaction is present also in the \Ros\ HRI data. As seen previously,
Fig.\,11 shows what appears to be a `bridge' of emission connecting the two galaxies. This type of
feature is believed to have been seen in several different situations, though importantly, 
always in association with galaxies in the process of interacting. For instance, a near-identical
feature is seen between the pair of interacting galaxies, NGC~3395/6 (Read \& Ponman
\cite{Read97P}), and at the contact region between the two galaxies making up the {\em Antennae}
system (Read \& Ponman \cite{Read95}). Furthermore, a similar feature is seen within the galaxy
group HCG~92 (Stefan's Quintet) (Pietsch \etal\ \cite{Pietsch97}). It is thought probable that all
of these features, including the present HRI feature between IC~4329A and IC~4329, are due to
shocks resulting from strong galaxy interactions.
So, though there is very strong evidence of an interaction occurring between IC~4329A and IC~4329,
whether, as regards the above one-sided blowout discussion, this interaction is strong enough to
displace the disc of IC~4329A with respect to the `central' wind source, is unclear.

There is another possible explanation of this very soft emission related to the harder
group emission discussed earlier. Remember that the high velocity dispersion of the
IC~4329A group galaxies and the high fitted temperature to the X-ray emitting gas
should point to the X-ray emitting gas having a high luminosity. This is observed
however, not to be the case, and it was suggested that `stripping' of some of the group
gas as the group moves through the surrounding A3574 cluster medium may be taking place.
This would result in a reduction in the X-ray luminosity and estimated mass of the
group gas, as we see. As discussed earlier, this idea is given some credence by the
fact that HCG~48, a group very similar to the IC~4329A group in terms of its high
velocity dispersion and high X-ray temperature, combined with a low X-ray luminosity
(Ponman \etal\ \cite{Ponman}), is also observed to be a group within a cluster (A1060),
and is thought to be falling through the centre of the cluster.

Could this soft emission, seen to the south-east of the IC~4329A/IC~4329 pair, be this
stripped group gas? The addition of the X-ray luminosity and gas mass of this soft
component to the equivalent values for the harder `group' component would certainly
push the total group gas $L_{X}$ and $M_{\mbox gas}$ to values such that the entire
properties of the IC~4329A group, taken as a whole, would appear completely normal and
typical. As mentioned in the introduction, this group lies at the easternmost edge of
the Hydra-Centaurus supercluster. Though no tangential velocities for these galaxies
are obviously available, it could perhaps be safe to assume that the IC~4329A group
would be moving tangentially in a general westward direction, towards the centre of
this superstructure. Any gas stripped from this group would appear to the east of the
group, as is observed in the case of the very soft emission.

Whether the temperature of this soft gas is consistent with this idea, is difficult to
say. One might have expected any stripped gas to be harder than its equivalent group
emission, though it might be possible that compression has lead to cooling. In fact,
the (roughly) estimated cooling time of this gas is fairly short (less than 3\,Gyr).

This idea, though interesting, may not be correct however, as perhaps the same
morphology should be seen in the case of the hard group emission. Though this hard
emission is severly contaminated by the IC~4329A wings, it does appear, on account of
the fact that the total hard emission is centred someway between IC~4329A and IC~4329,
that the hard group emission lies more predominantly to the opposite side of IC~4329A
than the soft emission.

\section{Summary}
\label{sec_summary}

We have observed both the ROSAT HRI and PSPC data from fields centred on the 
edge-on, type~1 Seyfert galaxy IC~4329A and its nearby companion, the giant 
lenticular IC~4329. 17 and 22 sources are detected respectively in the full HRI and 
PSPC fields of view, the brightest being associated with the two central galaxies 
and a further source to the south-west. Many coincidences are seen in the two
datasets and the nine most significant HRI sources all have equivalent PSPC
counterparts. In addition to point source emission, unresolved residual emission is
detected surrounding the IC~4329A/IC~4329 pair. This emission appears markedly
two-component, with a smooth, circularly-distributed hard component, centred midway
between the two central galaxies, and a more irregular, soft component, situated
almost entirely to the south-east of IC~4329A. Our findings with regard to the
observed point-source and unresolved emission can be summarized as follows:

1. An extremely bright ($L_{X}=6\times10^{43}$\,erg s$^{-1}$) source is detected
associated with the central Seyfert IC~4329A. Its spectral properties are compatible
with a single power-law ($\Gamma=1.73$), with a spectral break at 0.7\,keV, in
very good agreement with previous authors' work.

2. Two other very bright sources are detected associated with the nearby giant
lenticular IC~4329 ($L_{X}=8\times10^{41}$\,erg s$^{-1}$), and with a likely quasar to
the south-west. Fitting of standard spectral
models to these source spectra again result in fits that agree well with previous
authors' work.

3. Many other bright sources are detected both in the HRI and PSPC fields of
view, including three point-like sources, symmetrically positioned with
respect to the disc of IC~4329A. Optical follow-up observations of these sources
with the 2.2\,m ESO/MPG telescope at La Silla, Chile, has established that
they are nothing to do with the central Seyfert, being merely foreground and
background objects.
 
4. None of the sources detected show any significant temporal variability.

5. In addition to the point sources, residual emission is detected, both in the HRI
and in the PSPC, surrounding the IC~4329A/IC~4329 pair. This emission appears markedly
two-component, comprising of a hard, smooth, circularly-distributed component, 
centred somewhere between IC~4329A and IC~4329, and a soft, irregular component, 
situated almost entirely to the south-east of the IC~4329A disc. 

6. The hard component of the residual emission appears itself to be made up of two
components. One of these is purely the `wings' of the extremely bright IC~4329A
source, visible out to several arcminutes. The second component appears to be hot
($\sim$1.5\,keV) diffuse gas, with a luminosity of $\approx5\times10^{41}$\,erg
s$^{-1}$ and a mass of perhaps $2\times10^{11}$\,$M_{\odot}$. The properties of this
emission are very suggestive of it being due to hot gas within the galaxy group of
which IC~4329A and IC~4329 are members.

7. The soft component of the residual emission, in terms of its temperature and
one-sided nature, bears a good deal of resemblance to proposed starburst driven winds
seen in some far-infrared ultraluminous systems. It is however much brighter
($L_{X}=9\times10^{41}$\,erg s$^{-1}$), and larger. Another possibility discussed
briefly, is that the soft emission may be a `wake' of stripped gas from the galaxy
group.

8. A `bridge-like' feature is detected with the HRI between the two central galaxies, and 
is likely, as is seen in other similar systems, to be due to shocks resulting from the 
strong interaction taking place between the two systems.

\begin{acknowledgements}

AMR acknowledges the grateful receipt of a Royal Society fellowship during this work.
We are also very grateful to Andreas Vogler for his X-ray-optical overlay routines and
to Trevor Ponman for useful discussions. This research has made use of the SIMBAD
database operated at CDS, Strasbourg.

Optical images  are based on photographic data obtained with the UK Schmidt Telescope,
operated by the Royal Observatory Edinburgh, and funded by the UK Science and
Engineering Research Council, until June 1988, and thereafter by the Anglo-Australian
Observatory.  Original plate material is copyright (c) the Royal Observatory Edinburgh
and the Anglo-Australian Observatory. The plates were processed into the present
compressed digital form with their permission. The Digitized Sky Survey was produced at
the Space Telescope Science Institute under US Government grant NAG W-2166.

Finally, we thank our colleagues from the MPE ROSAT group for their support.
The ROSAT project is supported by the German Bundesministerium f\"ur Bildung, 
Wissenschaft, \\
Forschung und Technologie (BMBF/DARA) and the
Max-Planck-Gesellschaft (MPG).

\end{acknowledgements}

\end{document}